\documentclass[preprint, 12pt]{elsarticle}

\usepackage{amssymb}
\usepackage{subcaption}
\usepackage{graphicx}
\usepackage{verbatim}
\usepackage{mathtools}
\usepackage{xfrac}  
\usepackage{hyperref}
\usepackage{mathrsfs}
\usepackage{enumitem}
\usepackage{color}
\usepackage{booktabs}
\usepackage{natbib}
\usepackage[algoruled,boxed,lined]{algorithm2e}
\usepackage{algorithmic}
\usepackage[normalem]{ulem}
\usepackage{multirow}
\usepackage{tabularx}
\usepackage[labelformat=simple]{subcaption}

\def\bf#1{\boldsymbol{#1}}

\def\colr#1{\textcolor[rgb]{0.635,0.078,0.184}{#1}}  
\def\colo#1{\textcolor[rgb]{0.850,0.325,0.098}{#1}}

\def\colb#1{\textcolor[rgb]{0.000,0.4470,0.7410}{#1}}

\journal{arXiv}

\begin{document}
    \begin{frontmatter}
    \title{Optimal mesh generation for a non-iterative grid-converged solution of flow through a blade passage using deep reinforcement learning}
 
    \author{Innyoung Kim}
    \author{Jonghyun Chae}
    \author{Donghyun You \corref{cor1}}
    \ead{dhyou@postech.ac.kr}
    \address{Department of Mechanical Engineering, Pohang University of Science and Technology, 77 Cheongam-Ro, Nam-Gu, Pohang, Gyeongbuk 37673, South Korea\vspace{-0.4in}} 
		
    \cortext[cor1]{Corresponding author.}
	 
    \begin{abstract}
    An automatic mesh generation method for optimal computational fluid dynamics (CFD) analysis of a blade passage is developed using deep reinforcement learning (DRL). Unlike conventional automation techniques, which require repetitive tuning of meshing parameters for each new geometry and flow condition, the method developed herein trains a mesh generator to determine optimal parameters across varying configurations in a non-iterative manner. Initially, parameters controlling mesh shape are optimized to maximize geometric mesh quality, as measured by the ratio of determinants of Jacobian matrices and skewness. Subsequently, resolution-controlling parameters are optimized by incorporating CFD results. Multi-agent reinforcement learning is employed, enabling 256 agents to construct meshes and perform CFD analyses across randomly assigned flow configurations in parallel, aiming for maximum simulation accuracy and computational efficiency within a multi-objective optimization framework. After training, the mesh generator is capable of producing meshes that yield converged solutions at desired computational costs for new configurations in a single simulation, thereby eliminating the need for iterative CFD procedures for grid convergence. The robustness and effectiveness of the method are investigated across various blade passage configurations, accommodating a range of blade geometries, including high-pressure and low-pressure turbine blades, axial compressor blades, and impulse rotor blades. Furthermore, the method is capable of identifying the optimal mesh resolution for diverse flow conditions, including complex phenomena like boundary layers, shock waves, and flow separation. The optimality is confirmed by comparing the accuracy and the efficiency achieved in a single attempt with those from the conventional iterative optimization method.
    \end{abstract}
		
		
    \end{frontmatter}
	
    \section{Introduction} \label{sec1}
    A computational mesh is crucial for the accuracy, stability, and efficiency of computational fluid dynamics (CFD) simulations. However, generating such a mesh is a challenging and time-consuming endeavor that often necessitates extensive manual intervention. Despite geometric complexities, meshes with high orthogonality and smooth resolution variation are typically required to ensure accurate simulation results~\cite{ali2017optimal, zhang20182d}. This challenge is particularly pronounced in the CFD analysis of a blade passage in turbomachinery, where a structured mesh is preferred due to the wall-bounded nature of the flow, which exhibits specific directions~\cite{mueller2018stamps, marty2014numerical, kang2021low}. Nevertheless, creating a high-quality structured mesh that precisely conforms to the complex curvature of the blade while satisfying periodic boundary conditions in the azimuthal direction is not straightforward. Furthermore, even after the geometrical topology of the mesh is established, achieving an appropriate resolution to resolve critical flow features, such as near-wall boundary layers, presents a significant challenge, especially in the presence of shock waves and flow separation. Iterative simulations are often indispensable to identify the optimal resolution that effectively captures the crucial flow physics within available computational resources.
    
    To address the time-consuming and labor-intensive task of mesh generation, various mesh generators have been developed~\cite{milli2012padram, zagitov2014automatic, costenoble2022automated, marchandise2013cardiovascular, lu2020nnw, zhang20182d, zhu1991new, gargallo2018mesh, zheleznyakova2013molecular}. These tools automate the process, eliminating the need for manually specifying the location of individual nodes. However, due to the heuristic nature of most mesh generation processes, devising a universally applicable mesh generator solely through algorithmic definitions proves challenging. As a result, mesh generators typically require manual adjustment of certain parameters, such as the number of cells in each coordinate direction, expansion ratios, and clustering ratios. Engineers frequently need to fine-tune such meshing parameters for specific applications to improve mesh quality which deteriorates the reliability of simulation results depending on the expertise of the engineers.

    Efforts have been undertaken to establish a systematic approach to reduce human intervention in mesh generation. Optimization techniques, such as genetic algorithms and gradient-based methods, have been applied to mesh generation, yielding successful outcomes for various applications, including airfoils~\cite{dittmer2006mesh}, ground vehicles~\cite{ahmad2010mesh}, pressurized pipes~\cite{martins2014velocity}, and marine propellers~\cite{islam2015optimization}. This process entails evaluating the mesh based on geometrical qualities, such as orthogonality and skewness, or on the accuracy of the simulations, followed by adjustments of the meshing parameters. However, this approach inherently necessitates iterative cycles of mesh generation and evaluation to obtain optimal values. Furthermore, the optimal parameters identified are only valid to the configuration for which they were optimized. When faced with new geometries or flow conditions, the optimization process must be restarted from scratch, leading to a significant reduction in meshing efficiency.

    Deep reinforcement learning (DRL) has recently been applied to the field of mesh generation, leveraging its trial-and-error learning paradigm to automate the heuristic approaches typically required in mesh design. The capacity of DRL to dynamically adapt to varying state variables presents a promising solution to the generalization challenges inherent in conventional optimization techniques. Pan \textit{et al.}~\cite{pan2023reinforcement} successfully employed DRL to automatically generate high-quality quadrilateral meshes, identifying the capability of DRL to generalize across diverse geometrical configurations. However, the method lacked integration with simulation results, requiring additional adjustments to achieve the appropriate mesh resolution. For resolution identification, Foucart \textit{et al.}~\cite{foucart2023deep} applied DRL to optimize the adaptive mesh refinement (AMR) strategy for dynamic adjustment of the resolution to regions where needed. Yet, the iterative nature of AMR demands continuous error assessment and refinement cycles within the simulation, which could hinder the overall efficiency of the CFD workflow.

    The aim of the present study is to develop a method for producing an optimal mesh, both in terms of geometric quality and resolution, prior to simulation, thereby necessitating only a single simulation to attain optimal results for various geometries and flow conditions. The current study builds upon the work of Kim \textit{et al.}~\cite{kim2024non}, who utilized DRL to non-iteratively generate an optimal mesh for an arbitrary blade passage. In their approach, a mesh generator was developed and trained to optimally define meshing parameters to maximize geometric quality. Extending from the previous work, the CFD simulation results are integrated with the current study to optimize mesh resolution to fully eliminate human intervention. A wide range of blade profiles and flow conditions, including shock waves and flow separation, is considered with the Reynolds-averaged Navier--Stokes (RANS) simulation for versatile applicability in the industrial design of various fluid machinery such as gas and steam turbines and compressors. The objective is to yield an optimal CFD performance, balancing computational accuracy and cost in a multi-objective optimization framework. Thus, a converged solution can be acquired on the initial attempt, within available computational resources, thereby eliminating the iterative CFD processes typically required for grid convergence.
    
    However, the acquisition of training data by CFD simulations entails significant computational expense. To mitigate this, the training process is strategically divided into two steps, segmenting the dimension of action space (\emph{i.e.}, meshing parameters). In the first step, among the entire meshing parameters, those controlling the mesh shape are optimized to maximize geometric mesh quality following the previous work of Kim \textit{et al.}~\cite{kim2024non}. In the second step, the remaining resolution-related parameters are optimized by incorporating CFD results, using the mesh generator trained in the first step. The two-step strategy not only reduces the number of parameters requiring optimization through CFD but also enhances the stability and efficiency of computations, as the mesh used for the simulation is initially optimized for geometric quality. To further accelerate the data acquisition process, multi-agent reinforcement learning is employed, allowing multiple agents to conduct CFD simulations in parallel.

    The paper is organized as follows: Section~\ref{sec2} describes the theoretical background and the DRL algorithm employed in this study. Section~\ref{sec3} explains the computational frameworks for the current DRL training, including the CFD environment (Section~\ref{sec3.1}), the blade parameterization method (Section~\ref{sec3.2}), and the mesh generator (Section~\ref{sec3.3}). The state, action, and reward formulations and the learning procedures for the first and second training steps are detailed in Sections~\ref{sec4.1} and~\ref{sec4.2}, respectively. Section~\ref{sec5} investigates the effectiveness and robustness of the method by generating meshes for various blade passage configurations. Concluding remarks are provided in Section~\ref{sec6}.
    
    \section{Deep-reinforcement learning} \label{sec2}
    \subsection{Background} \label{sec2.1}
    DRL is a methodology for acquiring the optimal strategy in a complex decision-making process~\cite{sutton2018reinforcement}. In the context of the current study, DRL corresponds to a process of learning how to identify optimal meshing parameters given the geometries of the blades and the flow conditions. At each step $t$, a state $\bf{s}_{t} \in \mathcal{S}$ is provided and an action $\bf{a}_{t} \in \mathcal{A}$ is determined by a deterministic policy $\bf{\pi} : \mathcal{S} \rightarrow \mathcal{A}$, where $\mathcal{S}$ and $\mathcal{A}$ refer to state and action spaces, respectively. Upon executing the action $\bf{a}_{t}$, a reward $r_{t}(\bf{s}_{t}, \bf{a}_{t}) \in \mathbb{R}$ is received, and the subsequent state $\bf{s}_{t+1}$ is given following a probability distribution $p(\bf{s}_{t+1} \vert \bf{s}_{t}, \bf{a}_{t}) : \mathcal{S}\times\mathcal{A} \rightarrow \mathcal{P(\mathcal{S})}$, with $\mathcal{P}(\mathcal{S})$ representing the set of probability distributions over the state space $\mathcal{S}$. The procedure is iterated up to the terminal step $T$, at which point one episode concludes. The return $R_{t}$ is then computed as the summation of the immediate reward $r_{t}$ and the discounted future rewards as follows:
	\begin{equation}
        R_{t} = \sum_{i=t}^{T}\lambda^{(i-t)}~r_{i}(\bf{s}_{i}, \bf{a}_{i}),
        \label{eq_return}
    \end{equation}
    where $\lambda \in [0, 1]$ is a discount factor that controls the weight between short-term and long-term rewards.

    In DRL, a nonlinear neural network is employed to represent the policy $\bf{\pi}_{\bf{\phi}}$, where $\bf{\phi}$ corresponds to the network parameters, including weights and biases, which are adjusted during the training process. The aim is to identify an optimal policy $\bf{\pi}_{\bf{\phi}^{*}}$ that maximizes an objective function $J$ defined as the expectation of the return $R_{t}$. Accordingly, $\bf{\pi}_{\bf{\phi}}$ is updated by $\nabla_{\bf{\phi}}J$. This gradient is computed using the deterministic policy gradient algorithm~\cite{silver2014deterministic}, which employs the chain rule to differentiate $J$ as follows:
   \begin{equation}
    	\begin{aligned}
        \nabla_{\bf{\phi}}J & = {\mathop{\mathbb{E}}}_{\bf{s}_{t} \sim p}[\nabla_{\bf{\phi}}Q(\bf{s}_{t}, \bf{a}_{t})] \\
        & = {\mathop{\mathbb{E}}}_{\bf{s}_{t} \sim p}[\nabla_{\bf{a}} Q(\bf{s}_{t},\bf{a}_{t})|_{\bf{a}_{t}=\bf{\pi}_{\bf{\phi}}(\bf{s}_{t})}\nabla_{\bf{\phi}}\bf{\pi}_{\bf{\phi}}(\bf{s}_{t})],
        \label{eq_nabla_objfunction}
        \end{aligned}
    \end{equation}
    where $Q(\bf{s}_{t}, \bf{a}_{t})$ is an action value function defined as follows:
    \begin{equation}
        Q(\bf{s}_{t},\bf{a}_{t}) = {\mathop{\mathbb{E}}}_{\bf{s}_{i > t} \sim p}[R_{t}|\bf{s}_{t}, \bf{a}_{t}].
        \label{eq_Q}
    \end{equation}
    
    \subsection{Single-step deep-reinforcement learning} \label{sec2.2}
    As DRL is designed to achieve an optimal action at each discrete step $t$ within a sequential decision-making process, it exhibits outstanding performance in control problems where the current action directly influences the subsequent state. However, when addressing optimization problems, the primary goal is to precisely identify an optimal solution for a fixed problem definition. Once this optimal solution is discovered, the optimization task is considered complete; there is no subsequent problem to address. Accordingly, studies employing DRL to optimization problems often utilize a single-step DRL approach~\cite{vinquerat2021direct, ghraieb2022single, kim2022multi, viquerat2023policy, kim2024non}, wherein each learning episode consists of only a single step.
    
    In single-step DRL, each episode is completed once a reward $r(\bf{s}, \bf{a})$ is obtained following the execution of an action $\bf{a}$ in a given state $\bf{s}$, without transitioning to a subsequent state. Here, the subscript $t$ is omitted since the process involves only a single step. Consequently, both the return $R$ and the action value function $Q(\bf{s}, \bf{a})$ are reduced to the immediate reward $r$ as follows:
    \begin{equation}
        R = Q(\bf{s},\bf{a}) = r(\bf{s}, \bf{a}).
        \label{eq_single_step_return_Q}
    \end{equation}
    A direct consequence is that the learning proceeds to find the optimal policy $\bf{\pi}_{\bf{\phi}^{*}}(\bf{s})$, which maximizes the reward $r$, without the inclusion of additional discounted future rewards. This approach facilitates the precise identification of the optimal solution by defining the objective function as the reward. Furthermore, single-step DRL eliminates the need for gradual improvement of actions through sequential steps, determining the optimal action that maximizes the reward for a given state at once after sufficient training. In the context of the present study, this allows for the generation of optimal meshes as actions without the need for iterations, given various blade geometries and flow conditions defined as the state.

    \subsection{Single-step actor-critic algorithm} \label{sec2.3}
    In the present study, the actor-critic method~\cite{konda2000actor}, modified for single-step DRL, is employed. The method involves two types of neural networks: an actor, denoted as $\bf{\pi}_{\bf{\phi}}$, and a critic, denoted as $Q_{\bf{\zeta}}$. The network parameters $\bf{\phi}$ and $\bf{\zeta}$ represent the weights and the biases of the actor and critic networks, respectively. The actor network $\bf{\pi}_{\bf{\phi}} (\bf{s})$ is responsible for determining an optimal action that maximizes the reward in a given state, while the critic network $Q_{\bf{\zeta}} (\bf{s}, \bf{a})$ predicts the reward depending on the state and the action. To this end, the actor and critic networks are trained in each episode using loss functions $J_{\bf{\pi}}$ and $J_{Q}$, respectively, defined as follows:
    \begin{equation}
    J_{\bf{\pi}} = \frac{1}{N_b}\sum_{i=1}^{N_{b}} Q_{\bf{\zeta}}(\bf{s}_{i}, \bf{a}_{i}),
    \label{eq_actor_loss}
    \end{equation}  
    \begin{equation}
    J_{Q} = \frac{1}{N_b}\sum_{i=1}^{N_{b}} \left(r_{i} - Q_{\bf{\zeta}}\left(\bf{s}_{i}, \bf{a}_{i}\right)\right)^{2},
    \label{eq_critic_loss}
    \end{equation}
    where $N_b$ is the number of data set $(\bf{s}, \bf{a}, r)$ sampled from the buffer, where the data acquired are accumulated. The actor network is updated using the gradient ascent algorithm to achieve the policy that maximizes the reward as follows:
    \begin{equation}
    \bf{\phi} \gets \bf{\phi} + \alpha_{\bf{\pi}} \nabla_{\bf{\phi}}J_{\bf{\pi}},
    \label{eq_actor_update}
    \end{equation}
    where $\alpha_{\bf{\pi}}$ is the learning rate of the actor network. The deterministic policy gradient algorithm in Eq.~\eqref{eq_nabla_objfunction} is employed to calculate $\nabla_{\bf{\phi}}J_{\bf{\pi}}$ as follows:
    \begin{equation}
    \nabla_{\bf{\phi}}J_{\bf{\pi}} =  \frac{1}{N_b}\sum_{i=1}^{N_{b}} \left( \nabla_{\bf{a}_i} Q_{\bf{\zeta}}\left(\bf{s}_i,\bf{a}_i\right)|_{\bf{a}_i=\bf{\pi}_{\bf{\phi}}\left(\bf{s}_i\right)}\nabla_{\bf{\phi}}\bf{\pi}_{\bf{\phi}}\left(\bf{s}_i\right) \right),
    \label{eq_actor_deter_poli_grad}
    \end{equation}
    where $\nabla_{\bf{a}_i} Q_{\bf{\zeta}}\left(\bf{s}_i,\bf{a}_i\right)|_{\bf{a}_i=\bf{\pi}_{\bf{\phi}}\left(\bf{s}_i\right)}$ is obtained by the critic network. The critic network is updated using the gradient descent algorithm to minimize the error of reward prediction as follows:
    \begin{equation}
    \bf{\zeta} \gets \bf{\zeta} - \alpha_{Q} \nabla_{\bf{\zeta}}J_{Q},
    \label{eq_critic_update}
    \end{equation}
    where $\alpha_{Q}$ is the learning rate of the critic network and $\nabla_{\bf{\zeta}}J_{Q}$ is computed as follows:
    \begin{equation}
        \nabla_{\bf{\zeta}}J_{Q} = -\frac{2}{N_b}\sum_{i=1}^{N_{b}} \left(r_{i} - Q_{\bf{\zeta}}\left(\bf{s}_{i}, \bf{a}_{i}\right)\right)\nabla_{\bf{\zeta}}Q_{\bf{\zeta}}\left(\bf{s}_i,\bf{a}_i\right).
    \end{equation}

    \section{Computational framework} \label{sec3}
    \subsection{Computational fluid dynamics} \label{sec3.1}
    For the simulation of a blade passage, the open-source software SU2~\cite{economon2016su2} is employed to solve the two-dimensional steady-state compressible RANS equations, which are expressed as follows:
    \begin{equation}
        \nabla \cdot \bf{F}^{c} - \nabla \cdot \bf{F}^{v} = 0,
    \label{eq_comp_RANS}
    \end{equation}
    where $\bf{F}^{c}$ and $\bf{F}^{v}$ represent the convective and viscous fluxes, respectively, written as follows:
        \begin{equation}
        \bf{F}^{c} = 
        \begin{Bmatrix} 
        \rho \bf{v} \\
        \rho \bf{v} \otimes \bf{v} + \overline{\overline{I}}p \\
        \rho E \bf{v} + p \bf{v} 
        \end{Bmatrix},~
        \bf{F}^{v} = 
        \begin{Bmatrix} 
        0 \\
        \overline{\overline{\tau}} \\
        \overline{\overline{\tau}} \cdot \bf{v} + \kappa \nabla T
        \label{eq_conv_vis_flux}
        \end{Bmatrix},
    \end{equation}
    where $\rho$ is the fluid density, $\bf{v}$ is the velocity vector, $E$ is the total energy per unit mass, $\overline{\overline{I}}$ is the identity matrix, $p$ is the static pressure, $\kappa$ is the thermal conductivity, $T$ is the temperature, and $\overline{\overline{\tau}}$ is the viscous stress tensor defined as follows:
    \begin{equation}
        \overline{\overline{\tau}} = \mu \left(\nabla \bf{v} + \nabla \bf{v}^{T}\right) - \frac{2}{3}\mu \overline{\overline{I}} \left(\nabla \cdot \bf{v}\right),
    \label{eq_shear_stress_tensor}
    \end{equation}
    where $\mu$ is the viscosity.
    
    For turbulent flows, $\mu$ is divided into the dynamic viscosity $\mu_{d}$ and the turbulent viscosity $\mu_{t}$ based on the Boussinesq hypothesis~\cite{wilcox1998turbulence}, such that $\mu = \mu_{d} + \mu_{t}$. Similarly, the thermal conductivity is expressed as $\kappa = \kappa_{d} + \kappa_{t}$, where $\kappa$ is computed from the viscosity $\mu$, the specific heat capacity at constant pressure $c_{p}$, and the Prandtl number $Pr$, as $\kappa = (c_{p} \mu)/Pr$. The dynamic viscosity $\mu_{d}$ is calculated using the Sutherland's law~\cite{white2006viscous}. The turbulent viscosity $\mu_{t}$ is obtained from a turbulence model. In the present study, the shear stress transport–$k$–$\omega$ (SST-$k$-$\omega$)~\cite{menter1994two} turbulence model is employed, adopting the $k$-$\omega$ model near the wall and transitioning to the $k$-$\epsilon$ model in the free-stream using a blending function for accurate predictions in both the near-wall and core regions.
    
    To close Eq.~\eqref{eq_comp_RANS}, the equation of state for a perfect gas is employed with the specific heat ratio $\gamma$ and the gas constant $R$. The pressure is determined from $p = (\gamma - 1)\rho \left[ E - 0.5\left(\bf{v}\cdot\bf{v}\right) \right]$, the temperature is given by $T = p/(\rho R)$, and the specific heat capacity at constant pressure as $c_{p} = (\gamma R)/(\gamma - 1)$. The governing equations are discretized using a finite volume method (FVM). The convective fluxes are discretized using the centered Jameson-Schmidt-Turkel (JST) scheme~\cite{jameson1981numerical}. Gradients of the flow variables required to evaluate the viscous fluxes are calculated using a weighted least-squares method. 

\begin{figure}[]
    \centering
    \includegraphics[width=0.75\linewidth]{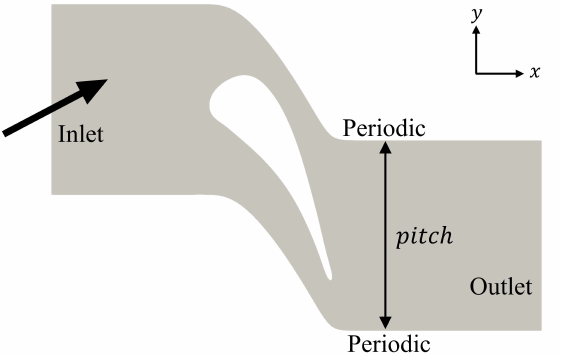}
    \caption{Computational configuration for a simulation of flow through a blade passage.} \label{fig_comp_config}
    \end{figure}

    The computational configuration is illustrated in Fig.~\ref{fig_comp_config}. At the inlet, the total pressure $p_{t,in}$ and the total temperature $T_{t, in}$ are imposed with inflow angle $\theta_{in}$. The static pressure $p_{out}$ is applied at the outlet. Periodic conditions are assigned for the upper and lower boundaries, which are separated by a distance of $pitch$. No-slip and adiabatic wall boundary conditions are prescribed at the blade surface.

    \subsection{Blade parametrization method} \label{sec3.2}
    For the generation of various blade types, including those in axial gas turbines, supersonic impulse turbines, and axial compressors, a parametrization method developed by Agromayor \textit{et al.}~\cite{AGROMAYOR2021} is utilized. A two-dimensional blade profile with a continuous curvature is produced using non-uniform rational basis spline curves~\cite{piegl1991nurbs} from blade shape parameters denoted as $\bf{BSP}$, written as follows:
    \begin{equation}
    \bf{BSP} = (C, \psi, \theta_{le}, \theta_{te}, d_{le}, d_{te}, \rho_{le}, \rho_{te}, t^{u}_{1}, \dots, t^{u}_{k}, t^{l}_{1}, \dots, t^{l}_{k}).
    \label{eq_BSP}\end{equation}

\begin{figure}[]
	\centering
	\begin{subfigure}[t]{0.49\linewidth}
	\includegraphics[width=1.0\linewidth]{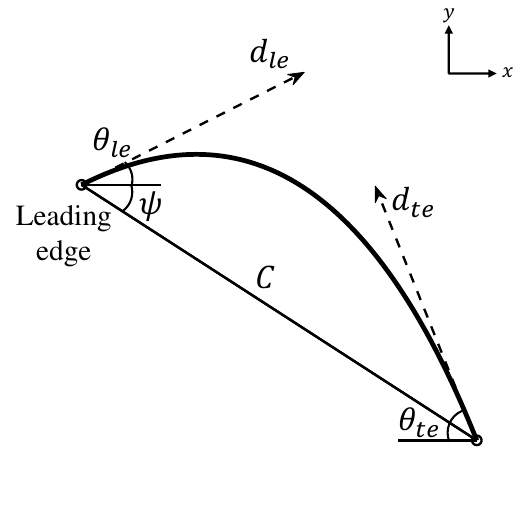}
	\caption{} \label{fig_parablade_a}
	\end{subfigure}
	\begin{subfigure}[t]{0.49\linewidth}
	\includegraphics[width=1.0\linewidth]{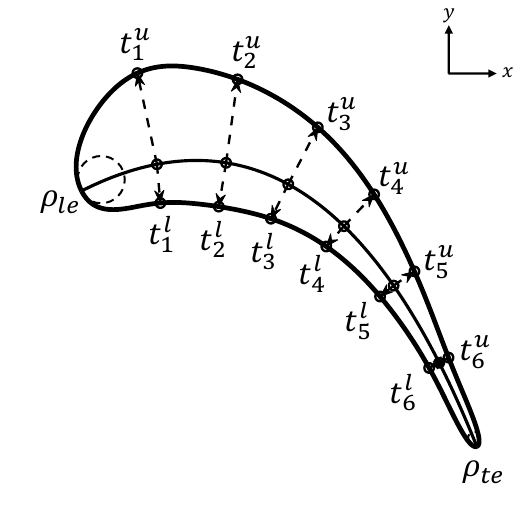}
	\caption{} \label{fig_parablade_b}
	\end{subfigure}
	\caption{Schematics of the blade parametrization method. (a) Camber line construction. (b) Blade profile construction.} \label{fig_parablade}
    \end{figure}

    The conceptual design of the method is depicted in Fig.~\ref{fig_parablade}. Initially, the camber line is established using the first six elements of $\bf{BSP}$, as illustrated in Fig.~\ref{fig_parablade_a}. The chord line of length $C$ is determined with the stagger angle $\psi$ from the leading edge. The camber line is then constructed with the curvature defined by the metal angles at the leading and trailing edges, denoted as $\theta_{le}$ and $\theta_{te}$, respectively. Here, the tangent proportions at the respective edges are controlled by $d_{le}$ and $d_{te}$. The elements that follow define the upper and lower profiles of the blade along the camber line, as illustrated in Fig.~\ref{fig_parablade_b}. The radii of curvature at the leading and trailing edges are determined by $\rho_{le}$ and $\rho_{te}$. The parameters $(t^{u}_{1}, \dots, t^{u}_{k})$ and $(t^{l}_{1}, \dots, t^{l}_{k})$ define the thickness distributions of the upper and lower sections of the blade, where $k$ indicates the number of parameters in the thickness distribution. For this study, $k = 6$ is chosen, as it has been found to be sufficient for accurately representing a diverse range of turbine blade profiles, aligning with the precision standards required in the manufacturing of blades for axial gas turbines~\cite{AGROMAYOR2021}.

\begin{figure}[]
    \centering
    \begin{subfigure}[t]{0.325\linewidth}
    \includegraphics[width=1.0\linewidth]{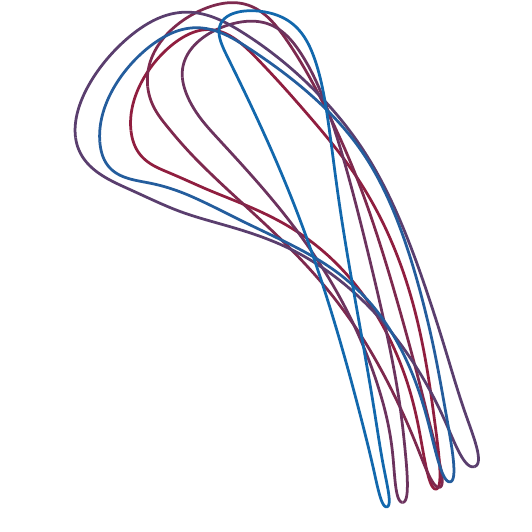}
    \caption{} \label{fig_blade_HPT}
    \end{subfigure}
    \begin{subfigure}[t]{0.325\linewidth}
    \includegraphics[width=1.0\linewidth]{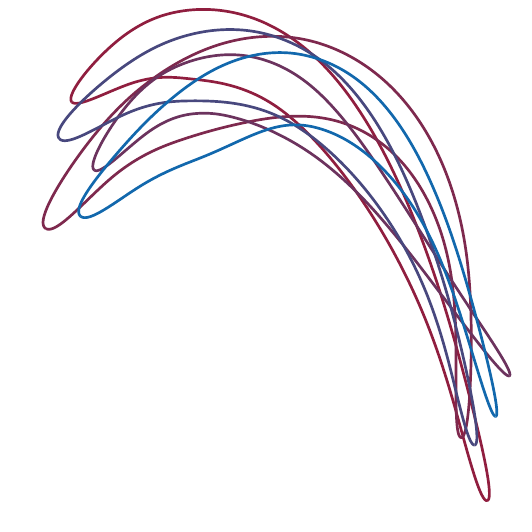}
    \caption{} \label{fig_blade_LPT}
    \end{subfigure}
    \begin{subfigure}[t]{0.325\linewidth}
    \includegraphics[width=1.0\linewidth]{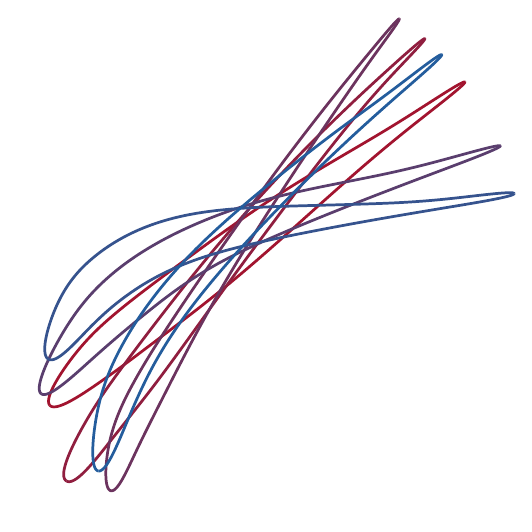}
    \caption{} \label{fig_blade_comp}
    \end{subfigure}
    \begin{subfigure}[t]{0.4\linewidth}
    \includegraphics[width=1.0\linewidth]{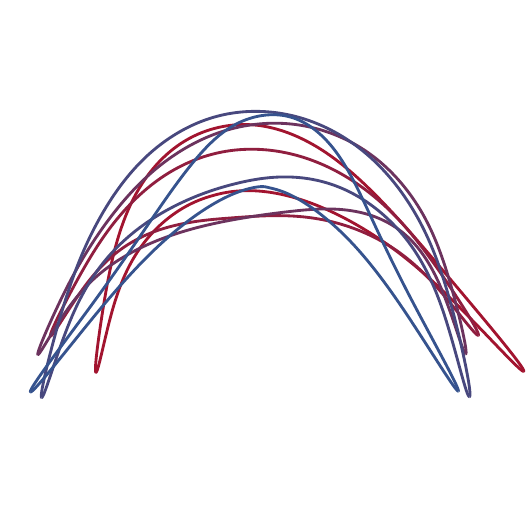}
    \caption{} \label{fig_blade_SIR}
    \end{subfigure}
    \begin{subfigure}[t]{0.4\linewidth}
    \includegraphics[width=1.0\linewidth]{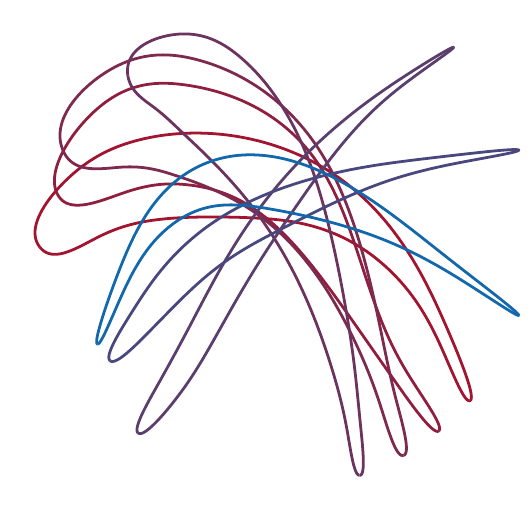}
    \caption{} \label{fig_blade_arb}
    \end{subfigure}
    \caption{Blade profiles produced using the parameterization method. Various blade geometries, including those for (a) high-pressure turbines, (b) low-pressure turbines, (c) axial compressors, (d) supersonic impulse turbines, and (e) blades uncommon in standard applications, can be generated. Each profile is distinguished by a unique color for clarity.} \label{fig_blade}
    \end{figure}

    The ranges of each element in $\bf{BSP}$ are set wide enough to include blades in the previous literature with different shapes and applications~\cite{arts1990aero, fransson1993panel, stadtmuller2001test, anand2020adjoint} as follows:
    \begin{equation}
        \centering
        \begin{array}{c}
        -90^{\circ} \leq \psi,~ \theta_{le},~ \theta_{te} \leq 90^{\circ}, \\
        0.4C \leq d_{le},~ d_{te} \leq 0.8C, \\
        0.0001C \leq \rho_{le},~ \rho_{te},~ t^{u}_{k},~ t^{l}_{k}  \leq 0.4C .\\        
        \end{array}
    \end{equation}\label{eq_BSP_range}
    Among these ranges, unrealistic blade shapes, such as those that are not simply connected or feature more than two extremal points, are excluded. Figure~\ref{fig_blade} depicts blade geometries attainable within the ranges. Diverse blade geometries, ranging from those with small to high camber and from blunt to sharp trailing edges, are produced with various angles of attack, including those for high-pressure turbines (Fig.~\ref{fig_blade_HPT}), low-pressure turbines (Fig.~\ref{fig_blade_LPT}), axial compressors (Fig.~\ref{fig_blade_comp}), supersonic impulse turbines (Fig.~\ref{fig_blade_SIR}), and even blades uncommon in standard applications (Fig.~\ref{fig_blade_arb}).

    \subsection{Mesh generator for a blade passage} \label{sec3.3}
    In the present study, the mesh generator that will be trained using DRL was developed by Kim \textit{et al.}~\cite{kim2024non}. The mesh generator produces a structured mesh for a two-dimensional blade passage using an elliptic mesh generation method~\cite{thompson1982elliptic, steger1979automatic, hsu1991numerical}. An HOH-type mesh, combining H-type meshes for the inlet and outlet sides with an O-type mesh near the blade, is produced by specifying meshing parameters. The descriptions of the parameters are listed in Table~\ref{table_params}.
	
\begin{figure}[]
    \centering
    \includegraphics[width=1.0\linewidth]{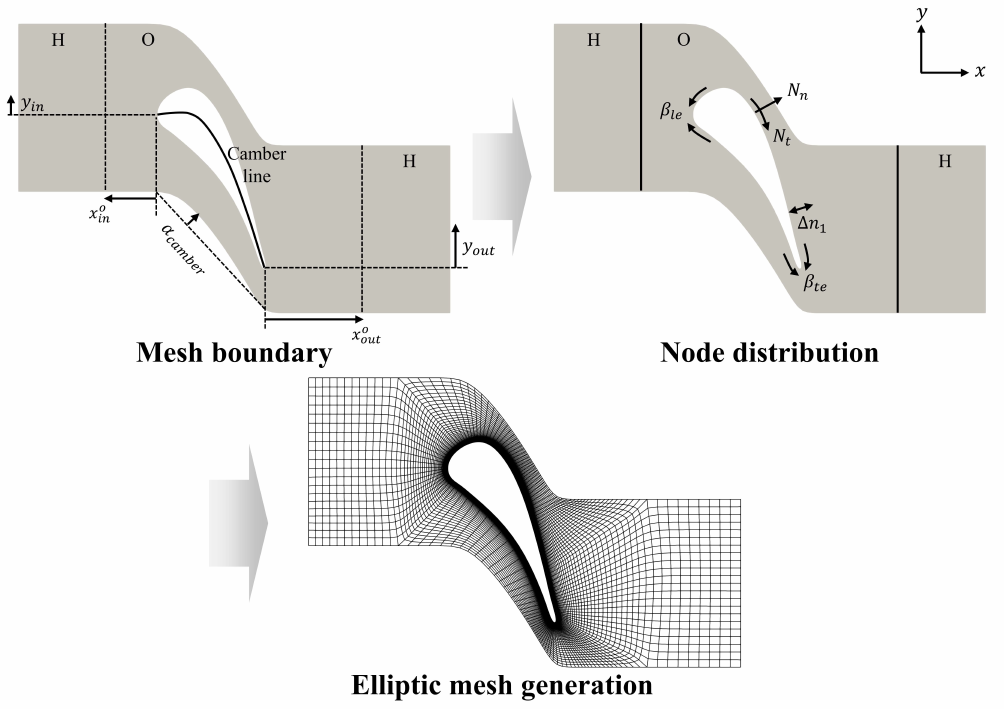}
    \caption{Schematic of the mesh generation algorithm. A structured mesh for a two-dimensional blade passage is produced by specifying meshing parameters that define the mesh boundary and the node distribution, followed by applying an elliptic mesh generation method.} \label{fig_mesh_generator}
    \end{figure}

    The schematic of the mesh generator is shown in Fig.~\ref{fig_mesh_generator}. Given the blade shape with $pitch$, the boundary shape of the mesh is defined based on the following parameters: $y_{in}$, $y_{out}$, $\alpha_{camber}$, $x^{o}_{in}$, and $x^{o}_{out}$. Here, $y_{in}$ and $y_{out}$ determine the positions of the inlet and the outlet in the vertical direction, respectively. To accommodate mesh generation for blades with varying curvatures, the curvature of the periodic boundaries is modulated by $\alpha_{camber}$, which ranges from a straight line $(\alpha_{camber} = 0)$ to fully conforming to the camber line $(\alpha_{camber} = 1)$. The locations of the interface between the O-type mesh and the H-type mesh at the inlet and outlet sides are determined by $x^{o}_{in}$ and $x^{o}_{out}$, respectively.

    Subsequently, the mesh resolution is determined based on $N_{t}$, $N_{n}$, $\beta_{le}$, $\beta_{te}$, and $\Delta n_{1}$. Initially, the numbers of nodes of the O-type mesh in the tangential and normal directions to the blade surface are specified by $N_{t}$ and $N_{n}$, respectively. Along the blade surface, $N_{t}$ nodes are distributed to ensure increased resolution at the leading and trailing edges. This distribution follows a hyperbolic tangent function, with $\beta_{le}$ and $\beta_{te}$ controlling the clustering intensity towards the leading and trailing edges, respectively. In the direction normal to the blade surface, $N_{n}$ nodes are distributed, clustering at the blade surface following a hyperbolic tangent function to adequately resolve near-wall physics. The clustering is defined such that the height of the first cell at the blade surface corresponds to $\Delta n_{1}$. After the nodes have been distributed within the domain, the mesh is ultimately generated using an elliptic mesh generation method~\cite{thompson1982elliptic, steger1979automatic, hsu1991numerical} to ensure high orthogonality near the blade surface without slope discontinuity.
    
    \section{Methodology} \label{4}
    \subsection{Geometric quality maximization} \label{sec4.1}
    \subsubsection{State, action, and reward} \label{sec4.1.1}
    The objective of the first-step training is to enable the mesh generator to autonomously select the geometry-controlling portions from the entire set of meshing parameters (see Table~\ref{table_params}) for various blade passage configurations. To achieve this, the state $\bf{s}$ is defined as:
    \begin{equation}
    \bf{s} = [\bf{BSP}, pitch, N_{o}, \beta_{le}, \beta_{te}, \Delta n_{1}],
    \label{eq_GM_state}
    \end{equation}
    where $\bf{BSP}$ denotes the blade shape parameters as outlined in Eq.~\eqref{eq_BSP}, $pitch$ refers to the blade spacing, and $N_{o} = N_{t} \times N_{n}$ represents the total number of nodes in the O-type mesh. The inclusion of $N_{o}$, $\beta_{le}$, $\beta_{te}$, and $\Delta n_{1}$ in the state, rather than as actions to be optimized, is intended to offer adaptability in mesh resolutions under various flow conditions. These parameters will be optimized in the subsequent training stage involving CFD simulations. 
    
    The action $\bf{a}$, which includes geometry-controlling meshing parameters, is defined as follows:
    \begin{equation}
    \bf{a} = [y_{in}, y_{out}, \alpha_{camber}, x^{o}_{in}, x^{o}_{out}, \delta],
    \label{eq_GM_action}
    \end{equation}
    where $\delta$ is incorporated to allocate the overall mesh resolution $N_{o}$, which is predetermined as a state variable, along the normal and tangential directions by $\delta = N_{t}/N_{n}$. 
    
    The reward is defined based on mesh quality, aiming to minimize numerical errors due to geometrical defects in the mesh, in accordance with the methodologies outlined in Kim \textit{et al.}~\cite{kim2024non}. To this end, two mesh quality metrics are considered: the ratio of determinants of the Jacobian matrices $\mathcal{Q}_{\mathcal{J}}$ and the cell skewness $\mathcal{Q}_{\mathcal{S}}$. The metrics $\mathcal{Q}_{\mathcal{J}}$ and $\mathcal{Q}_{\mathcal{S}}$ are crucial for achieving uniform cell distribution and maintaining high orthogonality between adjacent cells, respectively. Owing to these characteristics, the metrics are essential for generating a high-quality quadrilateral mesh, although the exact formulation might differ, as indicated in previous studies~\cite{zhang2006adaptive, knupp2000achieving, zhu1991new}.

    The calculation of $\mathcal{Q}_{\mathcal{J}}$ is as follows:
    \begin{equation}
    \mathcal{Q}_{\mathcal{J}} = \frac{\text{min}(\mathcal{J}_{i,j}, \mathcal{J}_{i+1,j}, \mathcal{J}_{i+1,j-1}, \mathcal{J}_{i,j-1})}{\text{max}(\mathcal{J}_{i,j}, \mathcal{J}_{i+1,j}, \mathcal{J}_{i+1,j-1}, \mathcal{J}_{i,j-1})},
    \label{eq_Qj}
    \end{equation}
    where, $\mathcal{J}_{i,j}$ is the determinant of the Jacobian matrix at each node $(x_{i, j}, y_{i,j})$, computed as: 
    \begin{equation}
	\mathcal{J}_{i,j} = 
	\begin{vmatrix}
	    \frac{\partial x}{\partial i} & \frac{\partial x}{\partial j} \\ 
        \frac{\partial y}{\partial i} & \frac{\partial y}{\partial j} 
	\end{vmatrix} = 
	\begin{vmatrix}
	    \dfrac{x_{i+1,j}-x_{i-1,j}}{2} & \dfrac{x_{i,j+1}-x_{i,j-1}}{2} \\ 
        \dfrac{y_{i+1,j}-y_{i-1,j}}{2} & \dfrac{y_{i,j+1}-y_{i,j-1}}{2} 
	\end{vmatrix}, 
	\label{eq_Jij}
    \end{equation}
    where $x$ and $y$ are the coordinates of the node, and $i$ and $j$ are the indices of the node. The value of $\mathcal{Q}_{\mathcal{J}}$ represents the area consistency among adjacent cells, indicating a smoother resolution change when the value is higher.

    The skewness $\mathcal{Q}_\mathcal{S}$ is calculated as follows:
    \begin{equation}
    \begin{gathered}
    \mathcal{Q}_\mathcal{S} = 1 - \text{max}\left(\frac{90^{\circ}-\theta_{\text{min}}}{90^{\circ}},\frac{\theta_{\text{max}}-90^{\circ}}{90^{\circ}}\right),
    \end{gathered}
    \end{equation}
    where $\theta_{\text{min}}$ and $\theta_{\text{max}}$ represent the minimum and maximum values among the interior angles of the cell. A higher value of $\mathcal{Q}_\mathcal{S}$ indicates reduced distortion of the cell and better orthogonality with neighboring cells.

    The reward $r$ is then defined as:
    \begin{equation}
    r = \left(\frac{(\mathcal{Q}_\mathcal{J})\vert_{min} + (\mathcal{Q}_\mathcal{J})\vert_{avg} + (\mathcal{Q}_\mathcal{S})\vert_{min} + (\mathcal{Q}_\mathcal{S})\vert_{avg}}{4}\right)^2,
    \label{eq_GM_reward}
    \end{equation}
    where $(~)\vert_{min}$ and $(~)\vert_{avg}$ represent the minimum and average values of the cells in the O-type mesh, respectively. This focus on the O-type mesh is due to the potential distortion of its cells to conform to the blade profile, as opposed to the cells of the H-type mesh, which typically remain rectangular. Extra nodes are incorporated at the boundaries of the O-type mesh for calculating $\mathcal{Q}_{\mathcal{J}}$, to ensure smooth transitions at mesh interfaces and periodic boundaries. By considering both the minimum and average values, the reward function is designed to effectively assess the impact of the cell with the lowest quality and to provide a comprehensive view of the overall quality distribution of the mesh. Note that the values of the metrics $(\mathcal{Q}_\mathcal{J})\vert_{min}$, $(\mathcal{Q}_\mathcal{J})\vert_{avg}$, $(\mathcal{Q}_\mathcal{S})\vert_{min}$, and $(\mathcal{Q}_\mathcal{S})\vert_{avg}$ range between 0 and 1 by definition, thereby exhibiting comparable scales. The square term is employed in the reward function to increase sensitivity to higher values.
    
    \subsubsection{Learning procedure} \label{sec4.1.2}
    The first-step training is conducted using the single-step actor-critic algorithm, described in Section~\ref{sec2.3}, with the detailed procedure presented in Algorithm~\ref{alg_drl_1st}. For each episode, a state $\bf{s}$ is randomly assigned. Based on the state, the actor determines an action as $\bf{a} = \bf{\pi}_{\bf{\phi}} (\bf{s})$ with exploration noise $\bf{\epsilon}$. The mesh generator then produces a mesh using the meshing parameters from the given state and the chosen action. Subsequently, the mesh quality of the generated mesh is evaluated, and the reward $r$ is calculated. The data of $(\bf{s}, \bf{a}, r)$ is then stored in a buffer. Utilizing the data from the buffer, the critic network is updated to more accurately predict the reward, while the actor network is refined to produce actions that maximize the reward, after which the next episode begins. This process is repeated until the convergence of the networks. Since the learning is conducted for a randomly given blade configuration each episode, rather than sequentially training on a single configuration and moving to the next, the converged network is capable of generating optimal meshes for a defined range of configurations.
    
    The actor and the critic are structured as fully-connected networks. All hidden layers utilize Leaky ReLU activation functions~\cite{maas2013rectifier}, with the exception of the output layer of the network, which employs a hyperbolic tangent function to bound the action values. The Adam optimizer~\cite{kingma2017adam} with a learning rate of $10^{-4}$ is employed to update the network parameters using a mini-batch size $N_b$ of $100$, a common approach in actor-critic algorithms~\cite{lillicrap2019, RN363}. For enhanced stability in the learning process, the actor network is updated every two updates of the critic network~\cite{RN363}. The exploration noise $\bf{\epsilon}$ is generated from a normal distribution $\mathcal{N}(0,\sigma^{2})$, with a mean of $0$ and the standard deviation $\sigma$, defined as follows:
    \begin{equation} \label{eq_expnoise}
    \sigma = \left\{ \begin{array}{ll}
    0.5\left\vert a_{max} - a_{min} \right\vert & \textrm{episode } \leq 1000, \\
    0.25 \left(\text{cos}\left(\frac{2\pi}{1000}\times\text{episode}\right)+1\right) \left\vert a_{max} - a_{min} \right\vert & \textrm{episode } > 1000. \\
    \end{array} \right.
    \end{equation}
    A higher $\sigma$ in the early episodes is used to gather a diverse set of data. Subsequently, a cosine function is employed to systematically vary the magnitude of the noise, balancing exploration and exploitation and reducing the influence of a specific noise value. Both networks comprise four hidden layers with $512$, $256$, $256$, and $128$ neurons, which have been identified to exhibit sufficient performance for the present case in the previous work of Kim \textit{et al.}~\cite{kim2024non}.

\begin{figure}[]
    \centering
    \centerline{\includegraphics[width=0.8\linewidth]{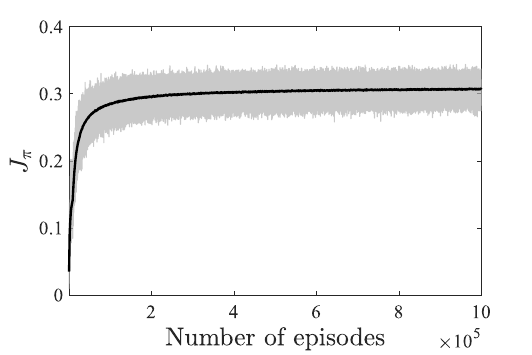}}
    \captionsetup{width=1.0\linewidth}
    \caption{The loss of the actor network $J_{\bf{\pi}}$, evaluated as a function of the number of episodes during the first-step training for geometric quality maximization. The grey line depicts the instantaneous loss, whereas the black line represents the moving average of the loss, computed over a span of $1000$ episodes.} \label{fig_first_step_conv}
    \end{figure}

    Figure~\ref{fig_first_step_conv} shows the loss of the actor network $J_{\bf{\pi}}$ in Eq.~\eqref{eq_actor_loss}, as a function of the number of episodes. As the number of episodes increases, the network is updated to produce meshes with higher geometric quality for each newly introduced blade passage configuration. Consequently, the accumulation of high-reward data in the buffer leads to an increase in the value of $J_{\bf{\pi}}$, which eventually converges. For the current optimization problem, approximately $10^{6}$ episodes are necessary to sufficiently train the network. The mesh generator trained in this phase will be utilized in the subsequent training phase, as detailed in Section~\ref{sec4.2}, with the aim of optimizing resolution-related meshing parameters through the integration of CFD simulations.

    \subsection{Training for optimal CFD} \label{sec4.2}
    \subsubsection{State and action formulation} \label{sec4.2.1}
    To fully eliminate human intervention in the mesh generation process, the remaining resolution-controlling parameters are optimized with CFD simulations and defined as the action $\bf{a}$ as follows:
    \begin{equation}
    \bf{a} = [N_{o}, \beta_{le}, \beta_{te}, \Delta n_{1}].
    \label{eq_CFD_train_action}
    \end{equation}

    The state $\bf{s}$ is formulated to reflect the computational configuration, including blade geometry and flow conditions, as follows:
    \begin{equation}
    \bf{s} = [\bf{BSP}, pitch, Re_{is, out}, Ma_{is, out}, \theta_{in}, w],
    \label{eq_CFD_train_state}
    \end{equation}
    where the blade geometry is captured by $\bf{BSP}$ and $pitch$ and the flow conditions are represented in a nondimensional manner by $Re_{is, out}$ and $Ma_{is, out}$, the isentropic Reynolds and Mach numbers at the outlet, respectively, along with the inflow angle $\theta_{in}$. The nondimensionalization is achieved using the boundary conditions ($p_{t, in}$, $T_{t, in}$, and $p_{out}$) and the axial chord length $C_{ax}$ of the blade as follows:
    \begin{equation}
    Ma_{is, out} = \sqrt{\frac{2}{\gamma - 1}\left( \left( \frac{p_{t, in}}{p_{out}}\right)^{\frac{\gamma - 1}{\gamma}} - 1 \right)},
    \label{eq_nondim_ma}
    \end{equation}
    \begin{equation}
    Re_{is, out} = \frac{\rho_{is, out} U_{is, out} C_{ax}}{\mu_{is, out}},
    \label{eq_nondim_re}
    \end{equation}
    \sloppy where $\rho_{is, out} = p_{out}/(RT_{is, out})$ by ideal gas law, $\mu_{is, out}$ is computed using the Sutherland's law~\cite{white2006viscous}, $U_{is, out} = Ma_{is, out}\sqrt{\gamma R T_{is, out}}$, and $T_{is, out} = T_{t, in} \left(p_{out}/p_{t, in}\right)^{\left(\gamma - 1\right)/\left(\gamma\right)}$.
    This nondimensionalization not only accommodates simulations of varying geometrical and physical scales but also reduces the data required for learning by consolidating the four variables ($p_{t, in}$, $T_{t, in}$, $p_{out}$, and $C_{ax}$) into two variables ($Re_{is, out}$ and $Ma_{is, out}$).
    
    The ranges of the flow conditions are defined wide enough to encompass complex flow features such as shock and separation, including those utilized in prior studies of blade passages~\cite{arts1990aero, mohanamuraly2021adjoint, sieverding2003turbine, kang2021low, stadtmuller2001test, ferrero2020field, zhao2020rans, michalek2012aerodynamic} as follows:
    \begin{equation}
    \centering
    \begin{array}{c}
    5\times10^4 \leq Re_{is, out} \leq 5\times10^6, \\
    0.4 \leq Ma_{is, out} \leq 1.1, \\
    \theta_{le} - 15^{\circ} \leq \theta_{in}  \leq \theta_{le} + 15^{\circ},        
    \end{array}
    \end{equation}\label{eq_flow_condition_range}
    where $\theta_{in}$ is adaptively adjusted based on the blade shape by defining the value relative to $\theta_{le}$, the metal angle at the leading edge, to mitigate unrealistic flow conditions.
    
    Finally, variable $w$ is incorporated as a state variable ranging from $0$ to $1$, not a fixed value, to determine the weight between two objectives: the cost and the error of the simulation. This approach allows users to obtain optimal meshes that balance accuracy and efficiency according to their preferences.

    \subsubsection{Designing the reward function} \label{sec4.2.2}
\begin{figure}[]
    \centering
    \centerline{\includegraphics[width=0.8\linewidth]{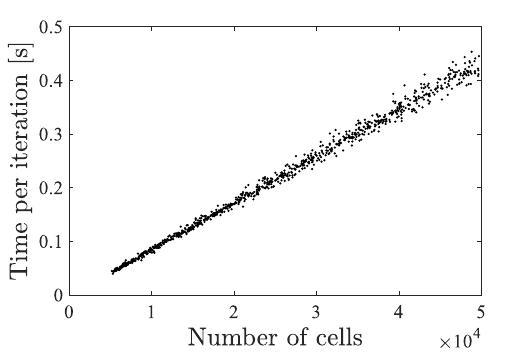}}
    \captionsetup{width=1.0\linewidth}
    \caption{Time taken per iteration for $1000$ randomly generated computational configurations as a function of the number of cells.} \label{fig_reward_cost}
    \end{figure}

    The reward function is devised to consider simultaneously the accuracy and the efficiency of the simulations. To achieve this, two objective functions are employed in a multi-objective approach: the cost and the error of the simulation. Firstly, the cost function, denoted as $r_{c}$, is designed to reflect the total time required for one simulation and is calculated as follows:
    \begin{equation}
    r_c = \text{iteration count} \times N_{total},
    \label{eq_CFD_train_reward_cost}
    \end{equation}
    where $N_{total}$ represents the total number of cells. This formulation is based on the observation that the time per iteration is $\mathcal{O}(N_{total})$, identified through simulations of $1000$ random flow configurations, as illustrated in Fig.~\ref{fig_reward_cost}. The linear scaling is attributed to the sparse matrix operation~\cite{barrett1994templates} in the current iterative solver, which utilizes the Flexible Generalized Minimum Residual (FGMRES) method~\cite{saad1993flexible} enhanced with an Incomplete Lower Upper (ILU) preconditioner~\cite{saad2003iterative}.
    
\begin{figure}[]
	\centering
	\begin{subfigure}[t]{0.6\linewidth}
	\includegraphics[width=1.0\linewidth]{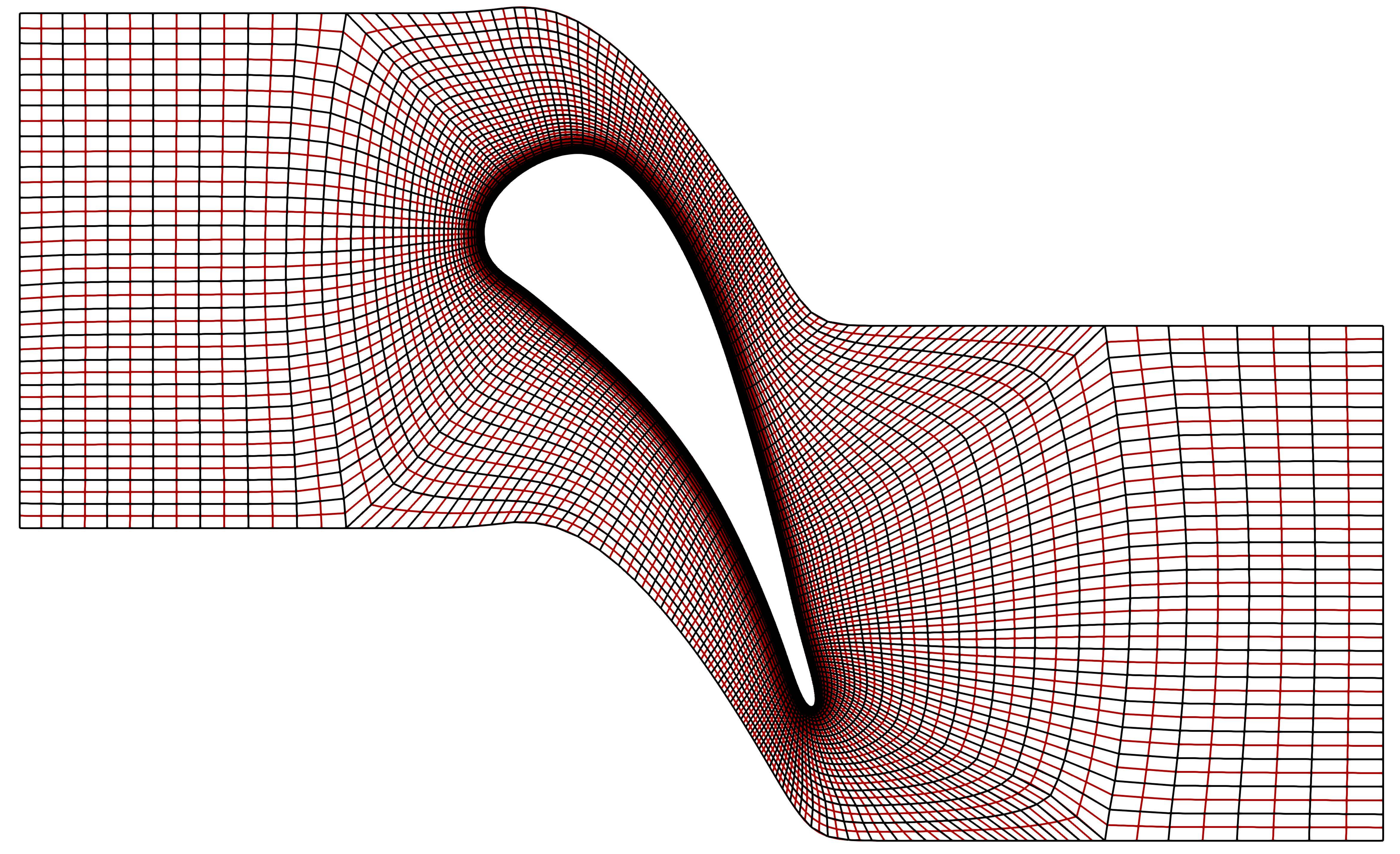}
	\caption{} \label{fig_reward_error_a}
	\end{subfigure}
	\begin{subfigure}[t]{0.39\linewidth}
	\includegraphics[width=1.0\linewidth]{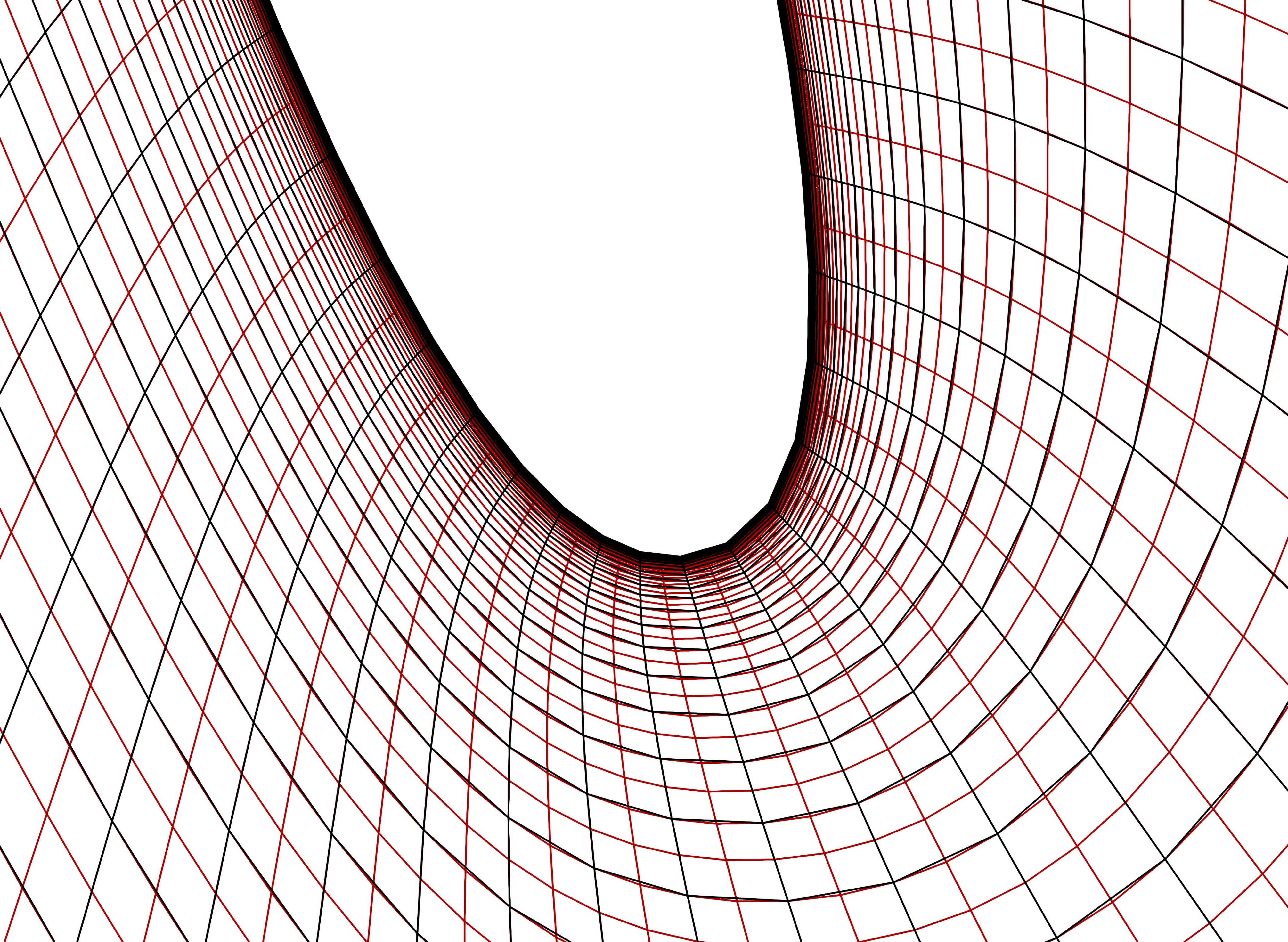}
	\caption{} \label{fig_reward_error_b}
	\end{subfigure}
	\caption{Meshes used to assess grid convergence. The black line represents the base mesh generated by the deep reinforcement learning (DRL) network. The red line represents the fine mesh, which is constructed by doubling the mesh resolution along each axis using Bicubic interpolation. (a) Full view of the mesh configuration. (b) Enlarged view around the trailing edge.} \label{fig_reward_error}
    \end{figure}

    The error function is designed to assess grid convergence by calculating the difference between simulation results obtained using two successively refined meshes: the base and fine meshes. The base mesh, initially produced by the DRL network, is further refined to create the fine mesh by doubling the resolution along each axis, as illustrated in Fig.~\ref{fig_reward_error_a}. Bicubic interpolation is employed to preserve the curvature of the blade, particularly at the leading and trailing edges, as depicted in Fig.~\ref{fig_reward_error_b}. The error function is defined by the root mean square (RMS) errors computed at the node locations of the base mesh, which remain unchanged during the refining process, as follows:
    \begin{equation}
    \begin{aligned}
    r_e &= \left. \text{RMS}\left(\left(Ma\right)_{fine} - \left(Ma\right)_{base}\right) \right\vert_{total~field} \\
    &+ \left. \text{RMS}\left(\left(Ma_{is}\right)_{fine} - \left(Ma_{is}\right)_{base}\right) \right\vert_{blade~surface}.
    \end{aligned}
    \label{eq_CFD_train_reward_error}
    \end{equation}
    Here, the error of $Ma$ is selected to capture essential flow features for blade passage analysis such as the boundary layer, the wake, and the shock wave~\cite{alauzet2022periodic}. Also, its nondimensional nature facilitates scaling across varied flow configurations. The error over the entire flow field is assessed by the first term, while the error along the blade surface is incorporated as the second term to reflect the near-wall flow physics. For the second term, the isentropic Mach number $Ma_{is}$ is computed by substituting the pressure along the blade surface for $p_{out}$ in Eq.~\eqref{eq_nondim_ma}.

    Finally, the reward function is defined using the weighted Chebyshev method, which guarantees the identification of all Pareto optimal solutions for both convex and nonconvex problems~\cite{miettinen2012nonlinear}, as follows: 
    \begin{equation}
    r = -\text{max}\left(w \times r_{e}^{*}, \left(1-w\right) \times r_{c}^{*} \right), 
    \label{eq_CFD_train_reward}
    \end{equation}
    where $w$ represents the weight between the two objective functions, which are scaled as follows:
    \begin{equation}
    \begin{aligned}    
    r_{e}^{*} &= \text{log}(r_{e}\times10^2 + 1), \\
    r_{c}^{*} &= \text{log}(r_{c}/10^{7} + 1).
    \end{aligned}
    \label{eq_CFD_train_reward_scale}
    \end{equation}
    Logarithmic scaling is employed to mitigate the impact of outliers during training. A factor of $10^2$ is multiplied by $r_e$, and $r_c$ is divided by $10^7$ to match the scales between the two objective functions for better learning performance.
    
\begin{figure}[]
    \centering
    \centerline{\includegraphics[width=0.6\linewidth]{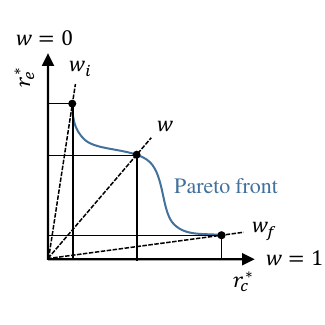}}
    \captionsetup{width=1.0\linewidth}
    \caption{Schematic illustration of the weighted Chebyshev method for determining Pareto front with two objective functions: the normalized error function $r_{e}^{*}$ and the normalized cost function $r_{c}^{*}$.} \label{fig_wcm}
    \end{figure}

    The schematic illustration of the weighted Chebyshev method for determining Pareto front is depicted in Fig.~\ref{fig_wcm}. For a given weight $w$, rectangles with a fixed width-to-height ratio and variable sizes are constructed on the plane of two objective functions, $r_{e}^{*}$ and $r_{c}^{*}$. As indicated by the minus sign in Eq.~\eqref{eq_CFD_train_reward}, the learning process aims to find a rectangle with the minimum size, resulting in a Pareto optimal solution. By varying the values of $w$ from 0 to 1, all Pareto front solutions can be obtained. Note that there exists an essential interval of weights (\emph{i.e.}, $[w_i, w_f] \subset [0, 1]$) for the generation of the entire front. As depicted in Fig.~\ref{fig_wcm}, every $w \in [0, w_i)$ and $w \in (w_f, 1]$ yields the same boundary solution of Pareto curve, which can be obtained using $w_i$ and $w_f$, respectively. The values of $w_i$ and $w_f$ can be determined by substituting each boundary solution into the condition where the vertex of the rectangle is formed in Eq.~\eqref{eq_CFD_train_reward}, as follows: 
    \begin{equation}
    \begin{aligned}    
    & w \times r_{e}^{*} = (1-w) \times r_{c}^{*}, \\
    & w = \frac{r_{c}^{*}}{r_{e}^{*} + r_{c}^{*}}.
    \end{aligned}
    \label{eq_CFD_train_reward_scale}
    \end{equation}

    To mitigate the high computational costs associated with acquiring data through CFD simulations, a method for data reproduction is devised to increase both the quantity and the diversity of data. Once $r_{e}$ and $r_{c}$ are obtained from a single simulation, various weights can be applied to these values to compute the reward function. For the present study, $100$ data sets are reproduced by randomly varying the weight $w$ within the range of $0$ to $1$ for each episode.
    
    \subsubsection{Learning procedure} \label{sec4.2.3}
\begin{figure}[]
    \centering
    \centerline{\includegraphics[width=1.0\linewidth]{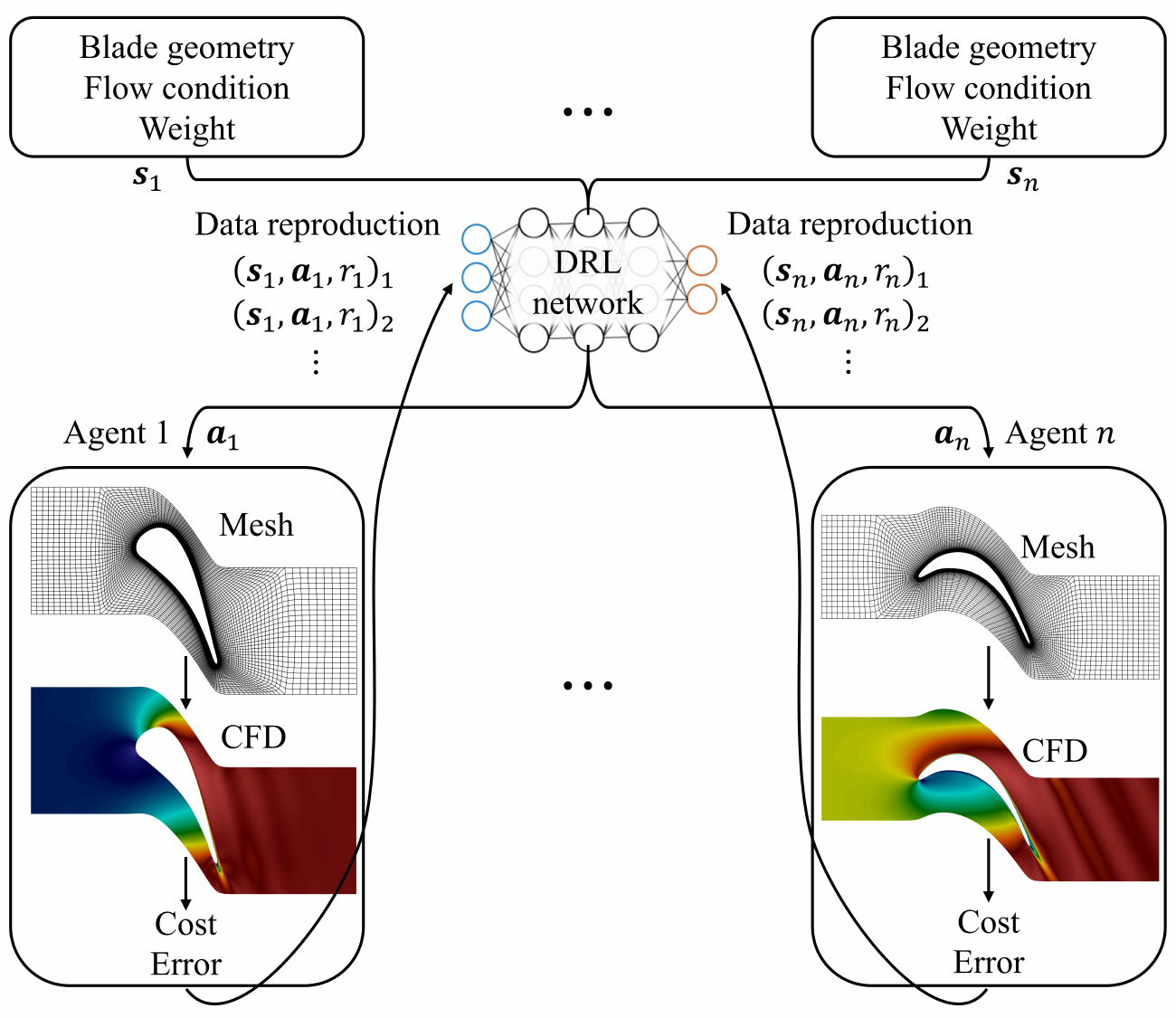}}
    \captionsetup{width=1.0\linewidth}
    \caption{Schematic of multi-agent reinforcement learning for automated mesh generation to achieve optimal computational fluid dynamics (CFD) simulations. States, each composed of the blade geometry, the flow condition, and the weight, are allocated to respective agents. Each agent then generates a mesh based on the action $\bf{a}$ designated by the DRL network and performs a CFD simulation. The simulation results are then used to determine the cost and the error, and the reward $r$ is computed, with the number of data being amplified using the data reproduction method.} \label{fig_MARL}
    \end{figure}

    Multi-agent reinforcement learning is employed to accelerate the data acquisition speed for training a mesh generator for optimal CFD. As illustrated in Fig.~\ref{fig_MARL}, multiple agents conduct CFD simulations in parallel to collect data, with the detailed procedure described in Algorithm~\ref{alg_drl_2nd}. States, each comprising the blade geometry denoted by $\bf{BSP}$ and $pitch$, the flow conditions indicated by $Re_{is, out}$, $Ma_{is, out}$, and $\theta_{in}$, and the weight $w$, are randomly assigned to individual agents. Each agent generates a mesh based on the action $\bf{a} = \bf{\pi}_{\bf{\phi}} (\bf{s})$ with exploration noise  $\bf{\epsilon}$, and then proceeds to perform a CFD simulation. If an agent $i$ completes the CFD simulation, the error and cost functions, $r_{e, i}$ and $r_{c, i}$, are calculated from the results, and the reward $r_{i}$ is computed. The number of data is subsequently amplified to $(\bf{s}_{i},\bf{a}_{i},r_{i})_{1}, \dots ,(\bf{s}_{i},\bf{a}_{i},r_{i})_{100}$ using the data reproduction method and stored in the buffer. The stored data is then used to update the actor and critic networks. Following this, the next episode begins, with a new task assigned to the corresponding agent. Note that tasks are dynamically reassigned as soon as agents complete the work, thereby increasing the data acquisition speed in proportion to the number of agents. This process is repeated until the networks converge. 
    
    The network architecture and hyperparameters are consistent with those established in the initial training phase, as described in Section~\ref{sec4.1.2}, with the exception that learning is conducted $100$ times per episode (\emph{i.e.}, $l_{max} = 100$) corresponding to the data amplification factor. Furthermore, the impact of network size is examined in \ref{app:network_size}, identifying that networks with four hidden layers provide adequate performance for the current optimization problem. 

\begin{figure}[]
    \centering
    \centerline{\includegraphics[width=0.8\linewidth]{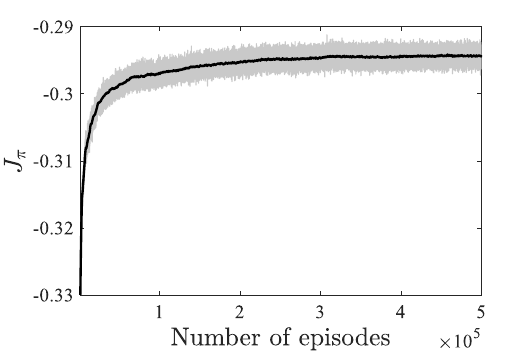}}
    \captionsetup{width=1.0\linewidth}
    \caption{The loss of the actor network $J_{\bf{\pi}}$, evaluated as a function of the number of episodes during the second-step training for optimal CFD. The grey line depicts the instantaneous loss, whereas the black line represents the moving average of the loss, computed over a span of $1000$ episodes.} \label{fig_second_step_conv}
    \end{figure}

    Figure~\ref{fig_second_step_conv} depicts the loss of the actor network $J_{\bf{\pi}}$ as a function of the number of episodes. A total of $5\times10^5$ RANS simulations are required to sufficiently train the network. The training employs $256$ agents over eight weeks, with each agent running on a single CPU core of an Intel(R) Xeon(R) E5-2650 v2 processor.
    
    \section{Results and discussion} \label{sec5}
    \subsection{Mesh generation for a high-pressure turbine blade} \label{sec5.1} 
    The capability of the trained network to generate optimal meshes is investigated for the LS89 turbine cascade, designed at the Von Karman Institute for Fluid Dynamics (VKI). The blade profile is representative of a high-pressure turbine nozzle guide vane found in modern aero-engines, for which experimental data are available in Arts et al.~\cite{arts1990aero}. Two test conditions are evaluated: MUR43, characterized by $Re_{is, out} = 5.90 \times 10^5$, $Ma_{is, out} = 0.84$, and $\theta_{in} = 0^{\circ}$, representing subsonic flow conditions; and MUR47, with $Re_{is, out} = 7.77 \times 10^5$, $Ma_{is, out} = 1.02$, and $\theta_{in} = 0^{\circ}$, indicating transonic flow conditions.

\begin{figure}[]
    \centering
    \begin{subfigure}[t]{0.43\linewidth}
    \includegraphics[width=1.0\linewidth]{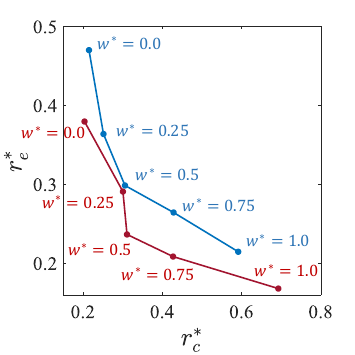}
    \caption{} \label{fig_result_HPT_pareto_a}
    \end{subfigure}
    \begin{subfigure}[t]{0.88\linewidth}
    \includegraphics[width=1.0\linewidth]{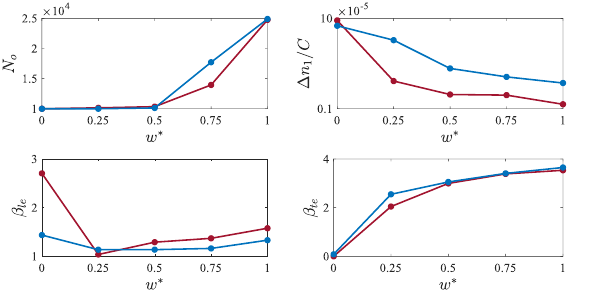}
    \caption{} \label{fig_result_HPT_pareto_b}
    \end{subfigure}
    \begin{subfigure}[t]{0.44\linewidth}
    \includegraphics[width=1.0\linewidth]{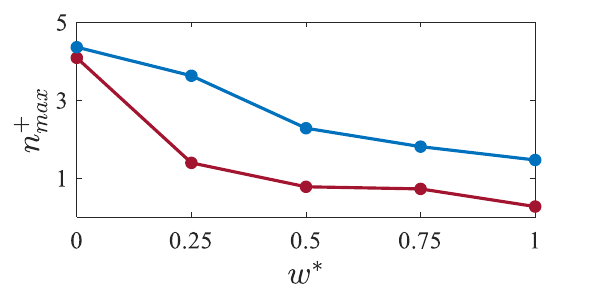}
    \caption{} \label{fig_result_HPT_pareto_c}
    \end{subfigure}
    \caption{(a) Pareto front obtained by the trained network for the LS89 cascade. (b) Corresponding optimal actions and (c) maximum values of dimensionless wall distance $n^{+}_{max}$ as a function of the normalized weight $w^{*}$. The red and blue colors correspond to the MUR43 and MUR47 test cases, respectively.} \label{fig_result_HPT_pareto}
    \end{figure}

\begin{figure}[]
    \centering
    \begin{subfigure}[t]{0.444\linewidth}
    \includegraphics[width=1.0\linewidth]{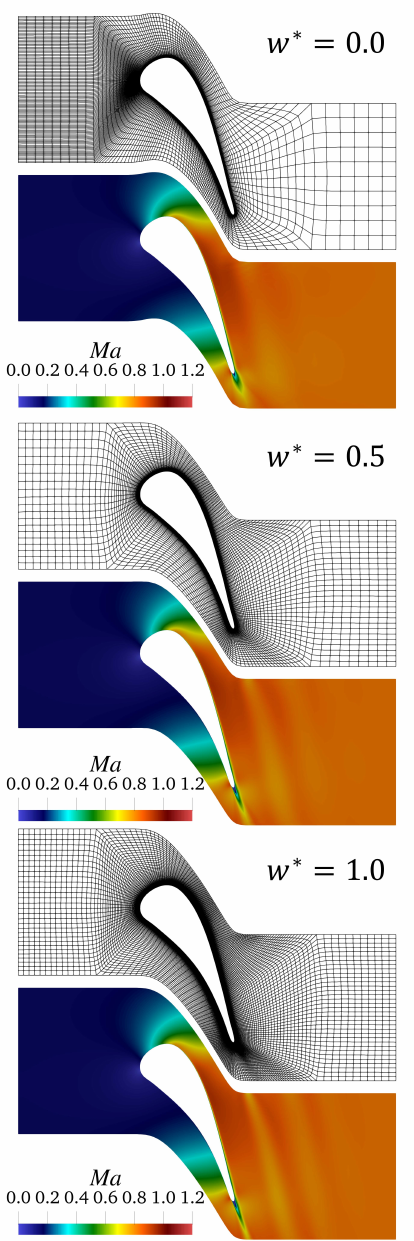}
    \caption{} \label{fig_HPT_sub}
    \end{subfigure}
    \begin{subfigure}[t]{0.444\linewidth}
    \includegraphics[width=1.0\linewidth]{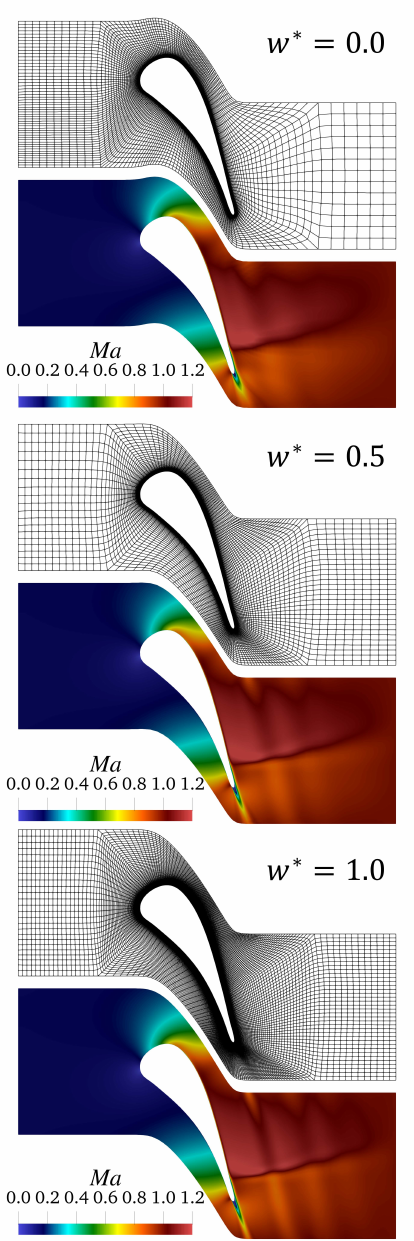}
    \caption{} \label{fig_HPT_super}
    \end{subfigure}
    \caption{Optimal meshes produced by the trained network and contour plots of the Mach number for (a) the MUR43 and (b) MUR47 test cases at $w^{*}$ of 0, 0.5, and 1.} \label{fig_HPT_mesh_ma}
    \end{figure}

    Figure~\ref{fig_result_HPT_pareto_a} shows Pareto front, and Figs.~\ref{fig_result_HPT_pareto_b} and \ref{fig_result_HPT_pareto_c} depict the corresponding optimal actions and the maximum values of the dimensionless wall distance $n^{+}_{max}$ as a function of the normalized weight $w^{*}$, respectively. The normalized weight is computed by linearly scaling the weight $w$ to its essential range, described in Section~\ref{sec4.2.2}, thereby enhancing the diversity of the solutions through increased weight variation. Corresponding meshes and contour plots of the Mach number are illustrated in Fig.~\ref{fig_HPT_mesh_ma}.

    As shown in Fig.~\ref{fig_result_HPT_pareto_a}, Pareto front is successfully obtained for both cases. The use of the weighted Chebyshev method enables the handling of nonconvexity, as seen in the MUR43 case at $w^{*} = 0.25$. The trade-off between the error function and the cost function is clearly observed. As depicted in Fig.~\ref{fig_result_HPT_pareto_b}, with increasing $w^{*}$, the network gradually increases $N_{o}$ from $10000$ to $25000$ and decreases $\Delta n_1$ to $n^{+}_{max} \approx 1$ (Fig.~\ref{fig_result_HPT_pareto_c}) to lower the error function. For $\beta_{le}$ and $\beta_{te}$, when $w^{*} = 0$, the nodes are excessively clustered toward the leading edge, resulting in insufficient resolutions in regions with large gradients of flow variables to achieve minimal cost function without consideration of the simulation error (see Fig.~\ref{fig_HPT_mesh_ma}). However, when the error function is incorporated into the objective function ($w^{*} > 0$), a more balanced mesh is achieved, characterized by a decrease in $\beta_{le}$ and an increase in $\beta_{te}$. As illustrated in Fig.~\ref{fig_HPT_mesh_ma}, the overall mesh resolution increases with higher $w^{*}$. This results in a more effective capture of the wake and yields a clearer definition of the shock location in the transonic case.

\begin{figure}[]
    \centering
    \begin{subfigure}[t]{0.495\linewidth}
    \includegraphics[width=1.0\linewidth]{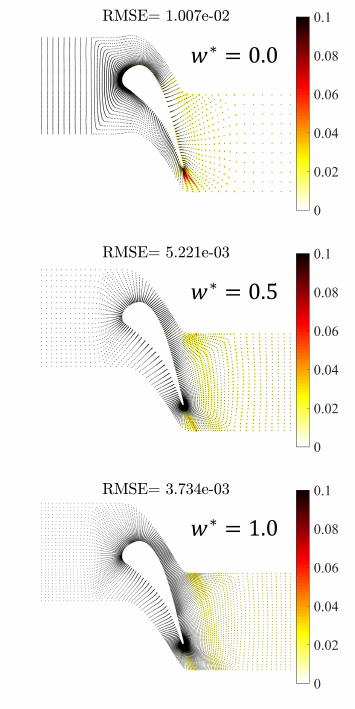}
    \caption{} \label{fig_HPT_ma_error_sub}
    \end{subfigure}
    \begin{subfigure}[t]{0.495\linewidth}
    \includegraphics[width=1.0\linewidth]{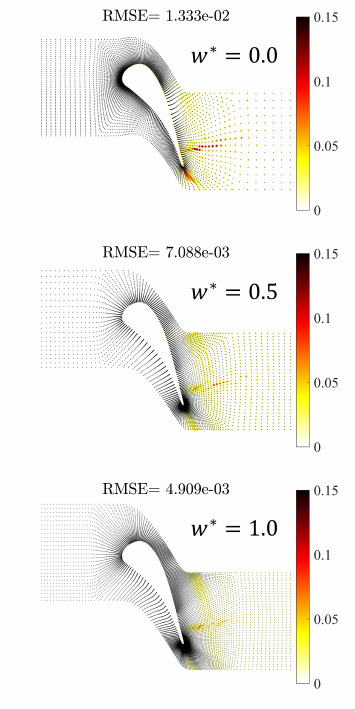}
    \caption{} \label{fig_HPT_ma_error_super}
    \end{subfigure}
    \caption{Error plots of the Mach number across the entire flow field ($\vert\left(Ma\right)_{fine} - \left(Ma\right)_{base}\vert$) and the corresponding root mean square error (RMSE) values for (a) the MUR43 and (b) MUR47 test cases at $w^{*}$ of 0, 0.5, and 1. The black dots indicate the node locations of the base mesh where the errors are computed.} \label{fig_HPT_ma_error}
    \end{figure}

    To further analyze the behavior of the network, the error functions for the corresponding optimal solutions are investigated. Figure~\ref{fig_HPT_ma_error} illustrates the first term in Eq.~\ref{eq_CFD_train_reward_error}, representing the error over the entire flow field. As shown in Fig.~\ref{fig_HPT_ma_error_sub}, for the MUR43 case at $w^{*} = 0$, large errors are observed in the rear part of the blade, including the wake, particularly near the trailing edge, resulting in an RMS value of $1.007\times10^{-2}$. This can be attributed to the excessive clustering of mesh resolution toward the leading edge, as previously discussed. With an increase in $w^{*}$, the mesh resolution is enhanced toward the trailing edge by decreasing $\beta_{le}$ and increasing $\beta_{te}$, as depicted in Fig.~\ref{fig_result_HPT_pareto_b}. Consequently, the errors decrease, with RMS values reducing to $5.221\times10^{-3}$ at $w^{*} = 0.5$ and to $3.734\times10^{-3}$ at $w^{*} = 1$. A similar trend is observed for the MUR47 case; however, significant errors are also noted near the shock location, resulting in higher RMS values compared to the MUR43 case, as illustrated in Fig.~\ref{fig_HPT_ma_error_super}. This distinction is also identified in the behavior of the network. To address the high error near the shock location, a denser mesh resolution is required. To this end, although $\beta_{te}$ exhibits similar values for both cases, the MUR47 case demonstrates smaller $\beta_{le}$, leading to an increased resolution toward the trailing edge (see Fig.~\ref{fig_result_HPT_pareto_b}). Additionally, while $n^{+}_{max}$ for the MUR43 case approaches 1 at $w^{*}$ values of 0.5 and 0.75 and subsequently decreases below 1 to minimize the error function, $n^{+}_{max}$ for the MUR47 case remains slightly above 1. This allows a limited number of nodes to be distributed away from the wall, enhancing the resolution in the wake region. Consequently, a higher resolution at the rear part of the blade is achieved in the MUR47 case compared to the MUR43 case with similar cell counts, as seen in Fig.~\ref{fig_HPT_mesh_ma}. The RMS values for the MUR47 case decrease from $1.333\times10^{-2}$ to $4.909\times10^{-3}$.

\begin{figure}[]
    \centering
    \begin{subfigure}[t]{0.495\linewidth}
    \includegraphics[width=1.0\linewidth]{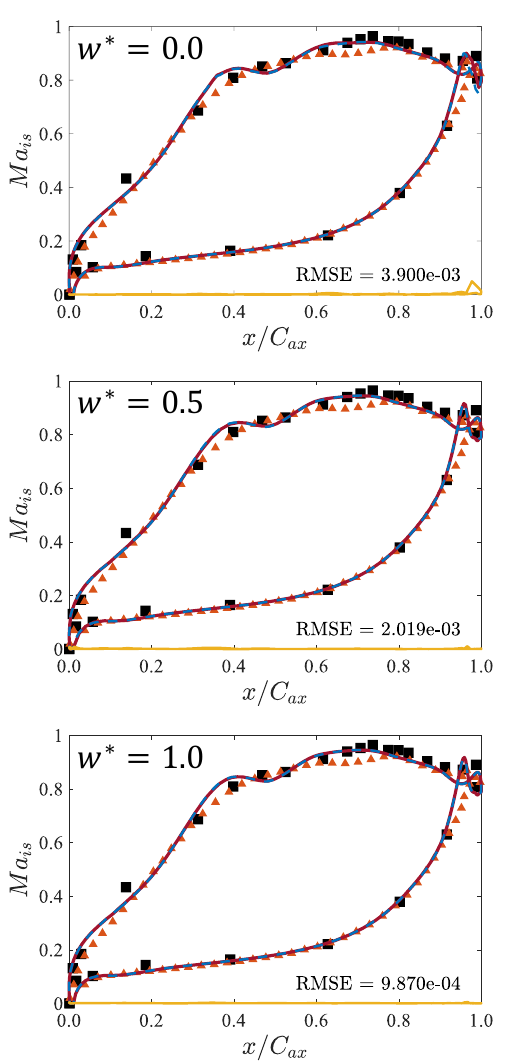}
    \caption{} \label{fig_HPT_mais_error_sub}
    \end{subfigure}
    \begin{subfigure}[t]{0.495\linewidth}
    \includegraphics[width=1.0\linewidth]{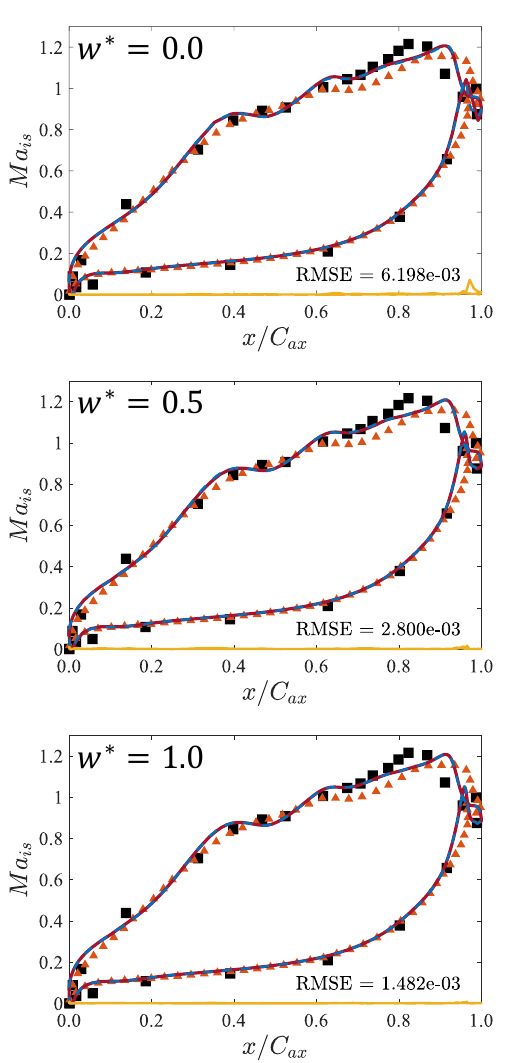}
    \caption{} \label{fig_HPT_mais_error_super}
    \end{subfigure}
    \caption{Isentropic Mach number along the blade surface for (a) the MUR43 and (b) MUR47 test cases at $w^{*}$ of 0, 0.5, and 1. The red solid lines and the blue dashed lines indicate results obtained from the base and fine meshes, respectively. The difference between the two meshes ($\vert\left(Ma_{is}\right)_{fine} - \left(Ma_{is}\right)_{base}\vert$) is depicted at the bottom as yellow solid lines, accompanied by the corresponding RMSE values. The black squares represent experimental data~\cite{arts1990aero}, while the orange triangles denote RANS simulation data of Mohanamural \textit{et al.}~\cite{mohanamuraly2021adjoint}.}\label{fig_HPT_mais_error}
    \end{figure}

    Figure~\ref{fig_HPT_mais_error} illustrates the second term of the error function from Eq.~\ref{eq_CFD_train_reward_error}, which represents the error along the blade surface. For both cases, the discrepancy in $Ma_{is}$ between the base and fine meshes diminishes as $w^{*}$ increases. The most significant reduction is observed near the trailing edge, which correlates with the earlier discussion on the enhancement of mesh resolution in this region. The RMS values decrease from $3.9\times10^{-3}$ to $9.87\times10^{-4}$ for the MUR43 case, and from $6.198\times10^{-3}$ to $1.482\times10^{-3}$ for the MUR47 case. Overall, the present results show good agreement with the experimental data, with the exception of the area following the shock. However, such discrepancies have also been reported in previous RANS simulation results~\cite{mohanamuraly2021adjoint, mueller2018stamps, li2023high}.
    
    \subsection{Mesh generation for a low-pressure turbine blade} \label{sec5.2}
    The capacity of the trained network is assessed for the T106C cascade, which is representative of low-pressure gas turbines in modern turbofan engines. The cascade was experimentally analyzed at the VKI and the results are available in the following literature~\cite{michalek2012aerodynamic, hillewaert2013dns}. In the current study, flow conditions with $Ma_{is, out} = 0.65$ and $\theta_{in} = 32.7^{\circ}$, along with two different $Re_{is, out}$ of $8\times10^4$ and $1.4\times10^5$, are investigated.

\begin{figure}[]
    \centering
    \begin{subfigure}[t]{0.43\linewidth}
    \includegraphics[width=1.0\linewidth]{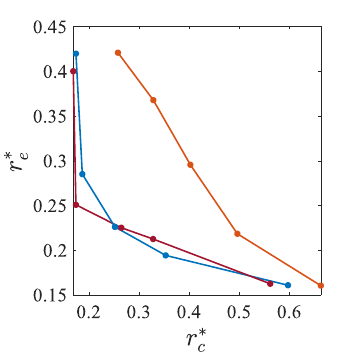}
    \caption{} \label{fig_result_LPT_pareto_a}
    \end{subfigure}
    \begin{subfigure}[t]{0.88\linewidth}
    \includegraphics[width=1.0\linewidth]{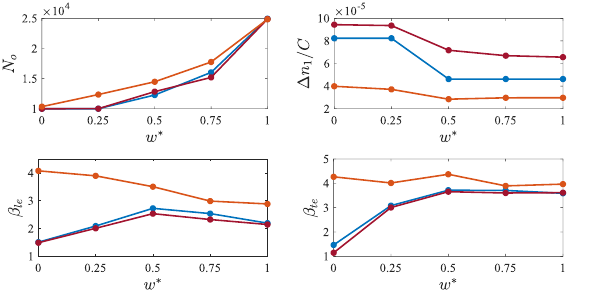}
    \caption{} \label{fig_result_LPT_pareto_b}
    \end{subfigure}
    \begin{subfigure}[t]{0.44\linewidth}
    \includegraphics[width=1.0\linewidth]{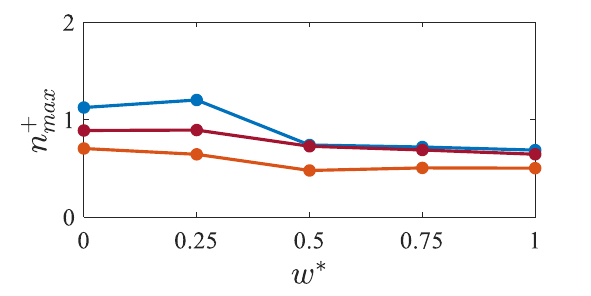}
    \caption{} \label{fig_result_LPT_pareto_c}
    \end{subfigure}
    \caption{(a) Pareto front obtained by the trained network for the T106C cascade. (b) Corresponding optimal actions and (c) $n^{+}_{max}$ as a function of $w^{*}$. The red and blue colors correspond to $Re_{is, out}$ of $8\times10^4$ and $1.4\times10^5$, respectively. The orange color indicate $Re_{is, out}$ of $1.4\times10^5$ with the transition model.} \label{fig_result_LPT_pareto}
    \end{figure}

\begin{figure}[]
    \centering
    \centerline{\includegraphics[width=0.9\linewidth]{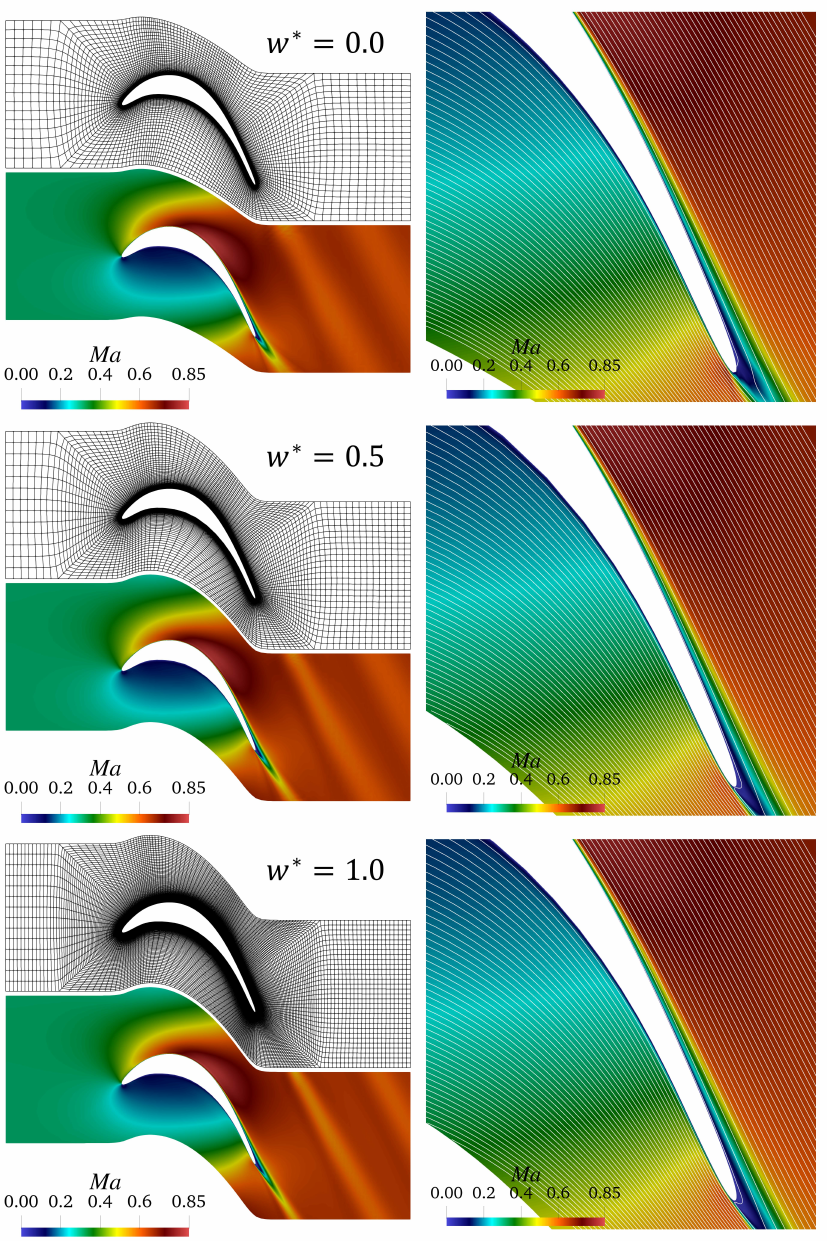}}
    \captionsetup{width=1.0\linewidth}
    \caption{Optimal meshes produced by the trained network and contour plots of the Mach number for $Re_{is, out}$ of $1.4\times10^5$ without the transition model at $w^{*}$ of 0, 0.5, and 1.} \label{fig_LPT_RE1_meshfield_mag}
    \end{figure}

    As shown in Fig.~\ref{fig_result_LPT_pareto_a}, Pareto front is successfully obtained, clearly illustrating the trade-offs between two objectives. In Figs.~\ref{fig_result_LPT_pareto_b} and \ref{fig_result_LPT_pareto_c}, for $Re_{is, out}$ of $1.4\times10^5$ (indicated by blue color), the network gradually increases $N_{o}$ and decreases $\Delta n_1$ to $n^{+}_{max}$ around 1 to reduce the error function. No significant changes are observed in $\beta_{le}$, while $\beta_{te}$ increases, thus enhancing the mesh resolution toward the trailing edge to more effectively capture the wake flow. This trend is also noted in the corresponding meshes and contour plots of the Mach number, as depicted in Fig.~\ref{fig_LPT_RE1_meshfield_mag}. Here, the curvature of the periodic boundaries is adjusted for different $w^*$ to achieve optimal geometric quality, as determined in the first-step training (Section~\ref{sec4.1}). 

\begin{figure}[]
    \centering
    \begin{subfigure}[t]{0.495\linewidth}
    \includegraphics[width=1.0\linewidth]{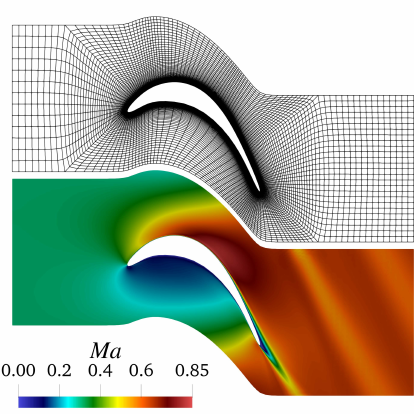}
    \caption{} \label{fig_LPT_0_meshfield}
    \end{subfigure}
    \begin{subfigure}[t]{0.495\linewidth}
    \includegraphics[width=1.0\linewidth]{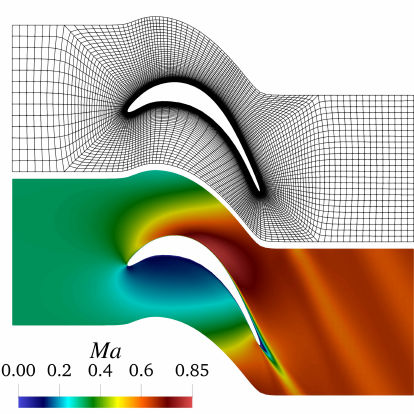}
    \caption{} \label{fig_LPT_1_meshfield}
    \end{subfigure}
    \begin{subfigure}[t]{0.495\linewidth}
    \includegraphics[width=1.0\linewidth]{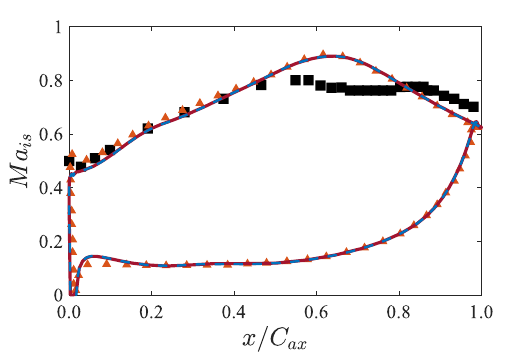}
    \caption{} \label{fig_LPT_0_mais}
    \end{subfigure}
    \begin{subfigure}[t]{0.495\linewidth}
    \includegraphics[width=1.0\linewidth]{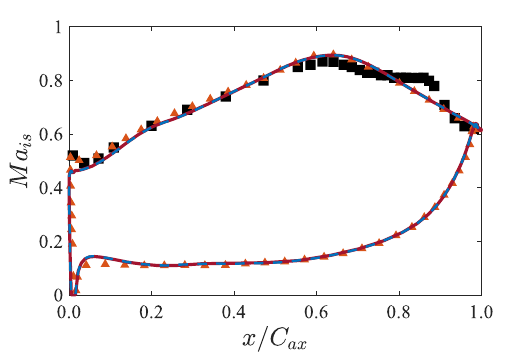}
    \caption{} \label{fig_LPT_1_mais}
    \end{subfigure}
    \caption{Optimal meshes constructed by the trained network and contour plots of the Mach number for $Re_{is, out}$ of (a) $8\times10^4$ and (b) $1.4\times10^5$, both without the transition model at $w^*$ of 0.5. The corresponding isentropic Mach number distribution along the blade surface is illustrated in (c) and (d), respectively. The red solid lines and the blue dashed lines indicate results obtained from the base and fine meshes, respectively. The black squares represent experimental data~\cite{michalek2012aerodynamic, hillewaert2013dns}, while the orange triangles denote RANS simulation data of Marty~\cite{marty2014numerical}.} \label{fig_LPT_notrans}
    \end{figure}

    For $Re_{is, out}$ of $8\times10^4$ (representd by red color), a higher $\Delta n_1$ is observed compared to $Re_{is_out}$ of $1.4\times10^5$. This is attributed to the formation of a thicker boundary layer at lower $Re_{is, out}$, where $n^{+}_{max}$ shows a similar scale around 1. While $n^{+}_{max}$ cannot be determined a priori and requires iterative simulations for near-wall resolution, the current method is highly efficient, producing results that satisfy the requirements in a single simulation. However, other actions do not exhibit notable differences between the two cases. This lack of distinction is due to the inability of the fully turbulent RANS model to predict flow separation. As indicated in Fig.~\ref{fig_LPT_RE1_meshfield_mag}, separation is not predicted for all $w^*$. Furthermore, as illustrated in Fig.~\ref{fig_LPT_notrans}, experimental data reveal flow separation characteristics, indicated by a plateau in $Ma_{is}$ distributions, differ for the two cases. Yet, the current simulations fail to capture the separation, resulting in similar outcomes for both cases. These limitations align with those reported in previous RANS simulation results~\cite {marty2014numerical}.

\begin{figure}[]
    \centering
    \centerline{\includegraphics[width=0.9\linewidth]{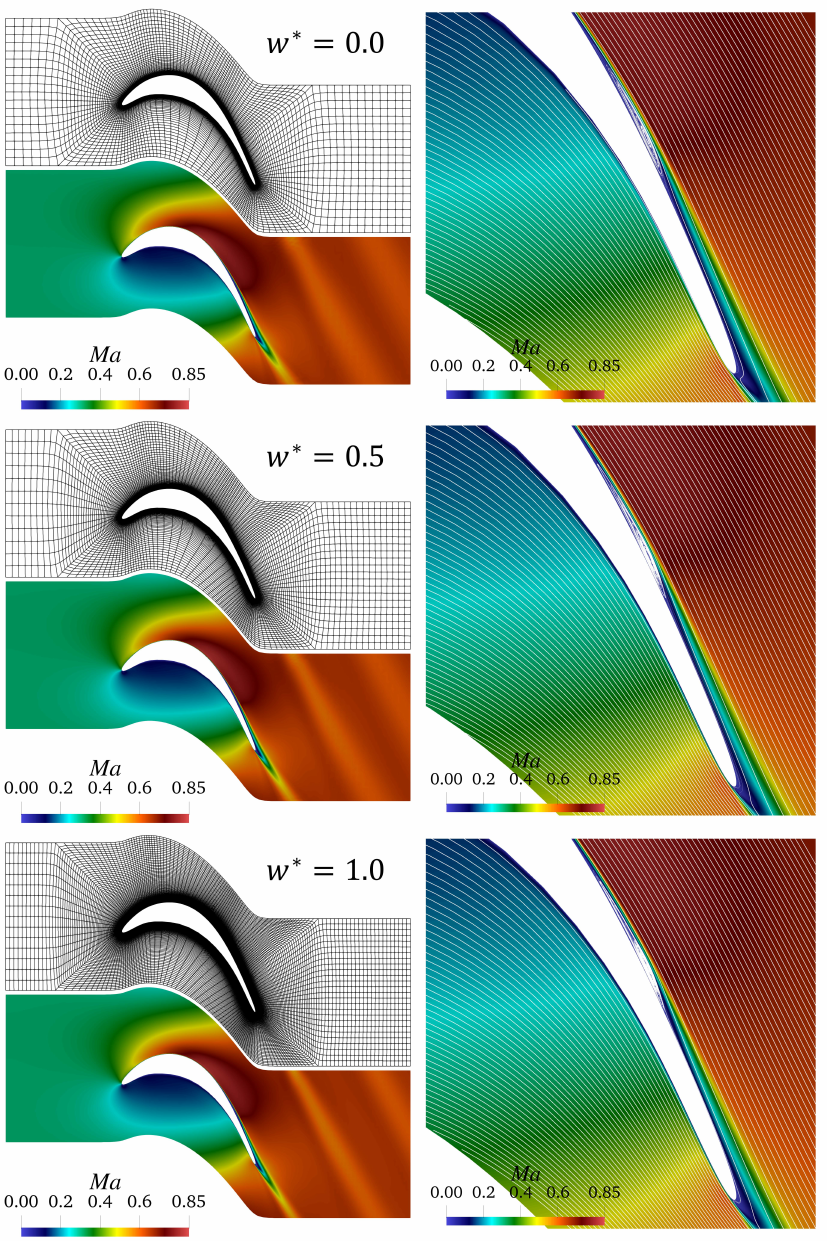}}
    \captionsetup{width=1.0\linewidth}
    \caption{Optimal meshes produced by the trained network and contour plots of the Mach number for $Re_{is, out}$ of $1.4\times10^5$ with the transition model at $w^{*}$ of 0, 0.5, and 1.} \label{fig_LPT_RE1trans_meshfield_mag}
    \end{figure}

    To investigate the impact of the flow separation on mesh optimization and to assess the applicability of the current method in flow conditions with the separation, additional training is conducted for $Re_{is, out}$ of $1.4\times10^5$ using the $\gamma-Re_{\theta}$ transition model~\cite{langtry2009correlation}. As represented by the orange color in Fig.~\ref{fig_result_LPT_pareto}, in comparison to the fully-turbulent case, higher values of $N_{o}$ are observed, while $\Delta n_{1}$ shows lower values with $n^{+}_{max} < 1$. Significant differences are observed in $\beta_{le}$ and $\beta_{te}$, which change node distribution along the blade surface, attributed to the predicted separation, as illustrated in Fig.~\ref{fig_LPT_RE1trans_meshfield_mag}. With an increase in $w^*$, both the location and extent of the separation are observed to converge. 

\begin{figure}[]
    \centering
    \begin{subfigure}[t]{0.495\linewidth}
    \includegraphics[width=1.0\linewidth]{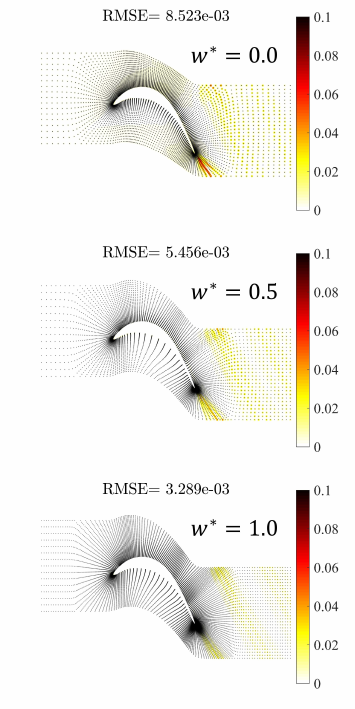}
    \caption{} \label{fig_LPT_ma_error_1}
    \end{subfigure}
    \begin{subfigure}[t]{0.495\linewidth}
    \includegraphics[width=1.0\linewidth]{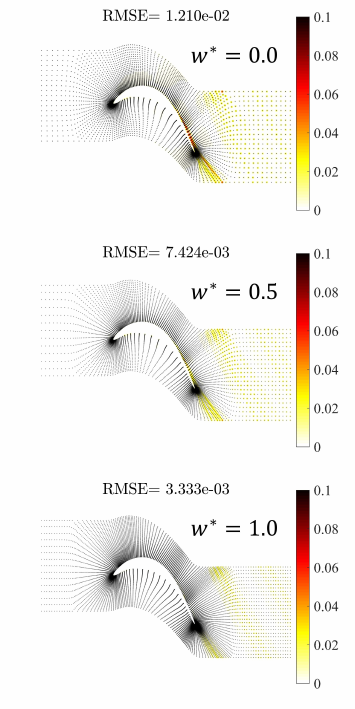}
    \caption{} \label{fig_LPT_ma_error_1trans}
    \end{subfigure}
    \caption{Error plots of the Mach number across the entire flow field ($\vert\left(Ma\right)_{fine} - \left(Ma\right)_{base}\vert$) and the corresponding root mean square error (RMSE) values for $Re_{is, out}$ of $1.4\times10^5$: (a) without and (b) with the transition model, at $w^{*}$ of 0, 0.5, and 1. The black dots indicate the node locations of the base mesh where the errors are computed.} \label{fig_LPT_ma_error}
    \end{figure}

    To further explore the behavior of the network in response to flow separation, error plots of the Mach number across the entire flow field are shown in Fig.~\ref{fig_LPT_ma_error}. In configurations without the transition model, significant errors are notable near the trailing edge and in the wake region. These errors diminish as the network increases mesh resolution toward the trailing edge, facilitated by an increase in $\beta_{te}$. This adjustment results in a reduction of RMS values from $8.523\times10^{-3}$ to $3.289\times10^{-3}$. With the transition model, considerable errors are observed in areas near the separation. To reduce these errors, the network decreases $\beta_{le}$, thereby enhancing resolution in the separation area. As a result, RMS values decrease from $1.21\times10^{-2}$ to $3.333\times10^{-3}$.

\begin{figure}[]
    \centering
    \begin{subfigure}[t]{0.495\linewidth}
    \includegraphics[width=1.0\linewidth]{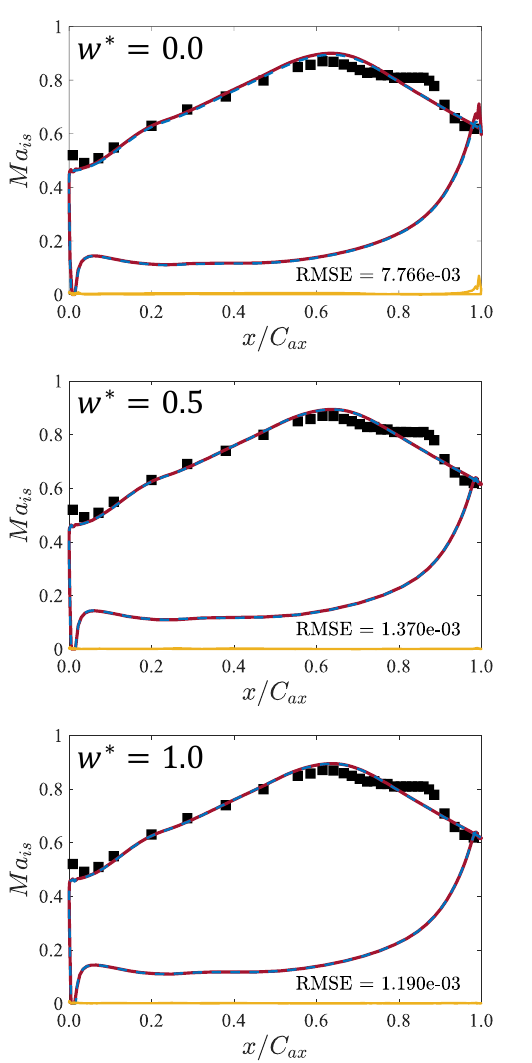}
    \caption{} \label{fig_LPT_mais_error_1}
    \end{subfigure}
    \begin{subfigure}[t]{0.495\linewidth}
    \includegraphics[width=1.0\linewidth]{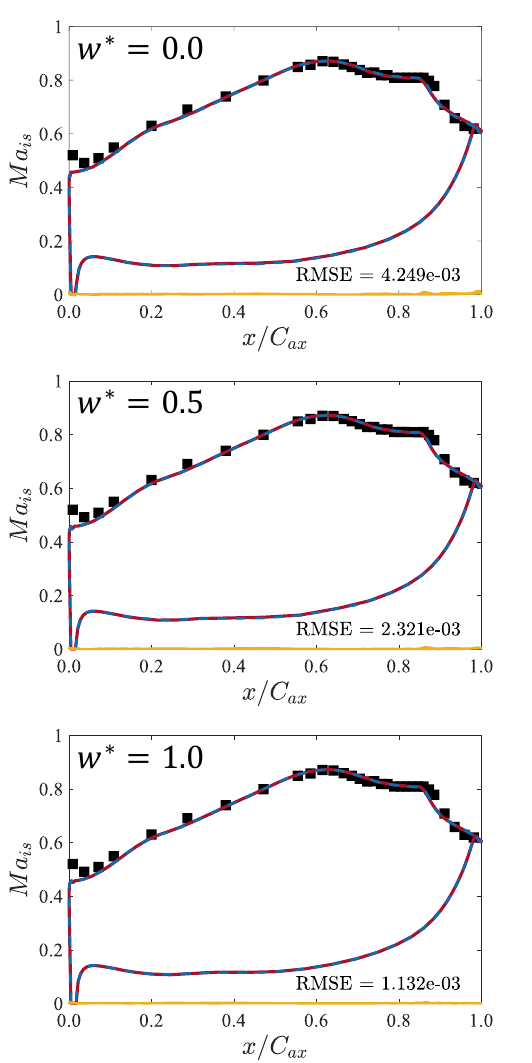}
    \caption{} \label{fig_LPT_mais_error_1trans}
    \end{subfigure}
    \caption{Isentropic Mach number along the blade surface for $Re_{is, out}$ of $1.4\times10^5$: (a) without and (b) with the transition model, at $w^{*}$ of 0, 0.5, and 1. The red solid lines and the blue dashed lines indicate results obtained from the base and fine meshes, respectively. The difference between the two meshes ($\vert\left(Ma_{is}\right)_{fine} - \left(Ma_{is}\right)_{base}\vert$) is depicted at the bottom as yellow solid lines, accompanied by the corresponding RMSE values. The black squares represent experimental data~\cite{michalek2012aerodynamic, hillewaert2013dns}.}\label{fig_LPT_mais_error}
    \end{figure}

    The distinctions in the error characteristics are also observed in $Ma_{is}$ distribution along the blade surface, as depicted in Fig.~\ref{fig_LPT_mais_error}. In both cases, the error decreases as $w^{*}$ increases. The results obtained without the transition model exhibit a notable reduction in error near the trailing edge, with RMS values declining from $7.766\times10^{-3}$ to $1.19\times10^{-3}$. However, with the transition model applied, the error reduction is more evenly distributed across areas experiencing separation, leading to a decrease in RMS values from $4.249\times10^{-3}$ to $1.132\times10^{-3}$. The results incorporating the transition model demonstrate good alignment with the experimental data.
    
    \subsection{Mesh generation for diverse blade passage configurations} \label{sec5.3}
\begin{figure}[]
    \centering
    \begin{subfigure}[t]{0.495\linewidth}
    \includegraphics[width=1.0\linewidth]{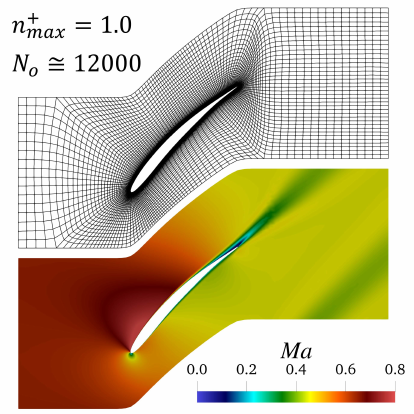}
    \caption{} \label{fig_arb_comp1_meshfield}
    \end{subfigure}
    \begin{subfigure}[t]{0.495\linewidth}
    \includegraphics[width=1.0\linewidth]{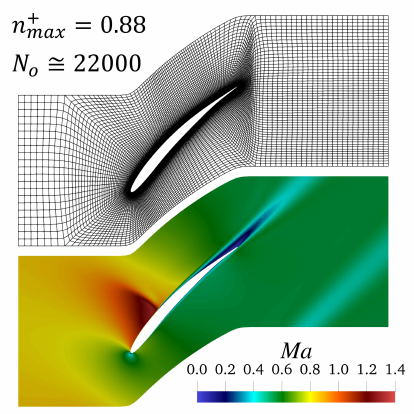}
    \caption{} \label{fig_arb_comp2_meshfield}
    \end{subfigure}
    \begin{subfigure}[t]{0.495\linewidth}
    \includegraphics[width=1.0\linewidth]{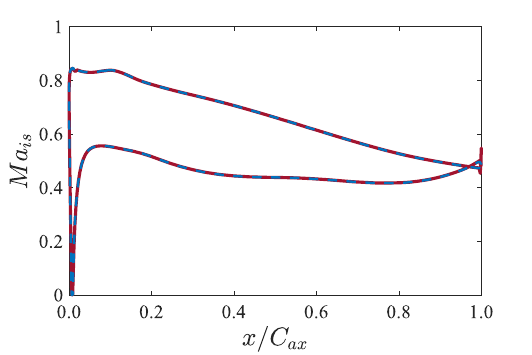}
    \caption{} \label{fig_arb_comp1_mais}
    \end{subfigure}
    \begin{subfigure}[t]{0.495\linewidth}
    \includegraphics[width=1.0\linewidth]{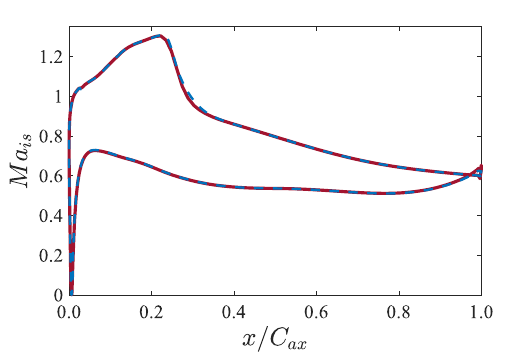}
    \caption{} \label{fig_arb_comp2_mais}
    \end{subfigure}
    \caption{Optimal meshes constructed by the trained network and contour plots of the Mach number for a compressor blade at $Ma_{is, out}$ of (a) $0.45$ and (b) $0.55$. The corresponding isentropic Mach number distribution along the blade surface is illustrated in (c) and (d), respectively. The red solid lines and the blue dashed lines indicate results obtained from the base and fine meshes, respectively.} \label{fig_arb_comp}
    \end{figure}

\begin{figure}[]
    \centering
    \begin{subfigure}[t]{0.495\linewidth}
    \includegraphics[width=1.0\linewidth]{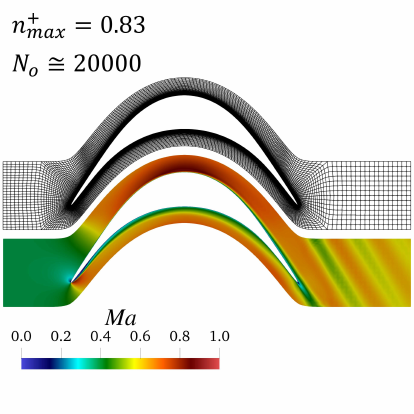}
    \caption{} \label{fig_arb_SIRT_meshfield}
    \end{subfigure}
    \begin{subfigure}[t]{0.495\linewidth}
    \includegraphics[width=1.0\linewidth]{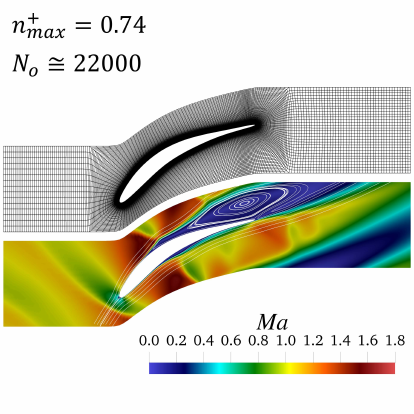}
    \caption{} \label{fig_arb_ext_meshfield}
    \end{subfigure}
    \begin{subfigure}[t]{0.495\linewidth}
    \includegraphics[width=1.0\linewidth]{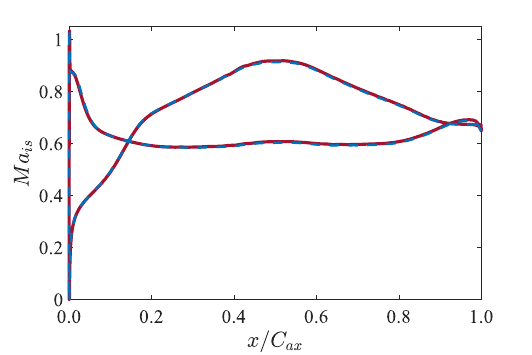}
    \caption{} \label{fig_arb_SIRT_mais}
    \end{subfigure}
    \begin{subfigure}[t]{0.495\linewidth}
    \includegraphics[width=1.0\linewidth]{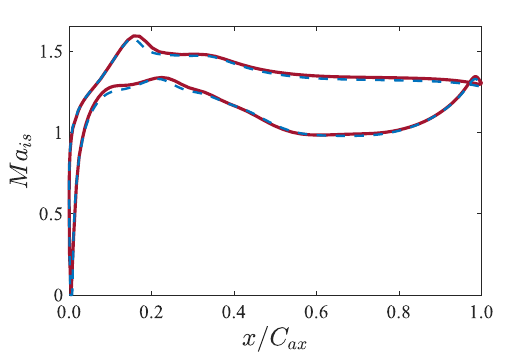}
    \caption{} \label{fig_arb_ext_mais}
    \end{subfigure}
    \caption{Optimal meshes generated by the trained network and contour plots of the Mach number for (a) an impulse turbine blade and (b) a blade operating under an extremely off-design condition. The corresponding isentropic Mach number distribution along the blade surface is illustrated in (c) and (d), respectively. The red solid lines and the blue dashed lines indicate results obtained from the base and fine meshes, respectively.} \label{fig_arb_2}
    \end{figure}

    To identify the versatility of the trained network, meshes for diverse configurations of blade passages are generated with a weight value of 0.5, as illustrated in Figs.~\ref{fig_arb_comp} and \ref{fig_arb_2}. The network achieves converged solutions comparable to those of fine meshes with $n^{+}_{max} \leq 1$ by merely specifying the flow configuration and the weight, thereby eliminating the iterative CFD processes typically required for grid convergence tests. Figure~\ref{fig_arb_comp} presents the results for an axial compressor blade under flow conditions with $Re_{is, out} = 5\times10^5$ and $\theta_{in} = 55^{\circ}$, and two distinct $Ma_{is, out}$ of 0.45 and 0.55. Note that a shock occurs even with a difference of 0.1 in $Ma_{is, out}$, necessitating an enhanced mesh resolution. Such situations are not known a priori; however, the current method is capable of handling such scenarios, ensuring an elevated resolution even with identical weights when a shock is present. As shown in Fig.~\ref{fig_arb_SIRT_meshfield}, the network is applicable to a blade for an impulse turbine, which is typically challenging to mesh due to high camber and sharp leading and trailing edges. Furthermore, the network effectively addresses a blade operating under an extremely off-design condition, where both separation and shock coexist, as shown in Fig.~\ref{fig_arb_ext_meshfield}.
 
    \subsection{Optimality analysis} \label{sec5.4}
    In this section, the optimality of the network, trained over a broad range of computational configurations, is examined. To this end, the normalized cost function $r_c^*$ and the normalized error function $r_e^*$, determined in a single attempt by the present network, are compared with those derived from conventional iterative optimization conducted on a single configuration. The LS89 cascade blade at the MUR43 and MUR47 test conditions, as discussed in Section~\ref{sec5.1}, is analyzed for $w^{*}$ values of 0, 0.5, and 1. The iterative optimization process follows the procedure described in Algorithm~\ref{alg_drl_2nd}, with the blade geometry, the flow condition, and the weight held constant. Since the weight is fixed, the data reproduction method is omitted. To mitigate the stochastic effects in DRL exploration and to enhance the probability of identifying the global optimum, iterative optimization is executed for five trials, each with different random seeds.
    
\begin{figure}[]
    \centering
    \begin{subfigure}[t]{0.495\linewidth}
    \includegraphics[width=1.0\linewidth]{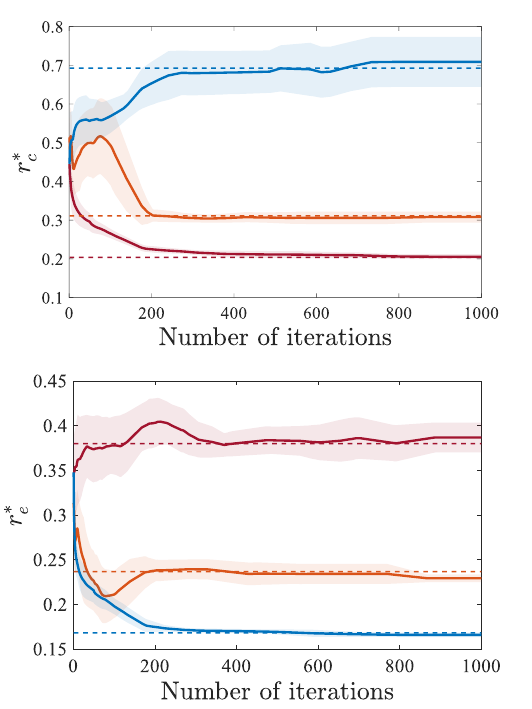}
    \caption{} \label{fig_optimality_sub}
    \end{subfigure}
    \begin{subfigure}[t]{0.495\linewidth}
    \includegraphics[width=1.0\linewidth]{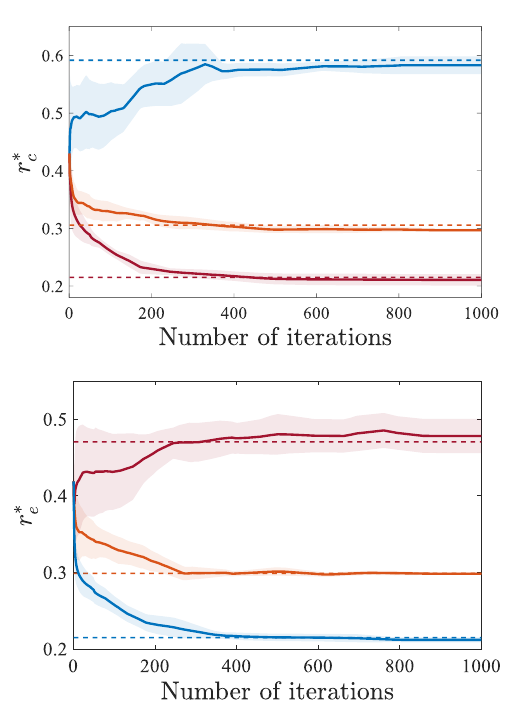}
    \caption{} \label{fig_optimality_super}
    \end{subfigure}
    \caption{The normalized cost function $r_c^*$ and the normalized error function $r_e^*$ as a function of the number of iterations for the LS89 cascade blade under (a) the MUR43 and (b) MUR47 test cases, obtained using iterative optimization from scratch. The bold lines represent the average of five runs with different random seeds, and the shaded areas indicate the standard deviation. The colors red, orange, and blue correspond to $w^{*}$ values of 0, 0.5, and 1. Additionally, the results obtained from a single attempt by the trained network are depicted with dashed lines for comparison.} \label{fig_optimality}
    \end{figure}

    Figure~\ref{fig_optimality} illustrates the results of the comparative analysis. The normalized cost function $r_c^*$ and the normalized error function $r_e^*$ are monitored as a function of the number of iterations. In both the MUR43 and MUR47 cases, the average values at $w^{*} = 0$ converge to the lowest $r_c^*$ and the highest $r_e^*$, while the average values at $w^{*} = 1$ demonstrate the opposite trend, with $w^{*} = 0.5$ representing intermediate values. Convergence of these average values is observed by the $10^3$th iteration, with the standard deviation diminishing, indicating consistent optimization results across different random seeds. Exceptions are noted in the secondary objective function when optimization is directed only towards a primary objective, such as with $r_e^*$ at $w^{*} = 0$ and $r_c^*$ at $w^{*} = 1$. In all cases, the present network achieves at least 96\% of the performance compared to iterative optimization. Notably, the current network demonstrates enhanced performance in the secondary objective for $w^{*}$ values of 0 and 1 in the MUR43 case and for $w^{*}$ values of 0 in the MUR47 case. This enhanced performance is attributed to the network being trained across a full range of weights, enabling it to learn the relationship between objective functions, in contrast to iterative optimization, which solely focuses on the primary objective.
    
    \section{Concluding remarks} \label{sec6}
    An automatic mesh generation method for the optimal computational fluid dynamics (CFD) analysis of a blade passage has been developed using deep reinforcement learning (DRL). A single-step DRL approach has been employed to train a mesh generator, enabling autonomous adjustment of meshing parameters to their optimal values in a non-iterative manner. During each training episode, the mesh generator creates a mesh and conducts CFD analysis for each randomly assigned flow configuration. The training has been conducted until the highest levels of simulation accuracy and computational efficiency have been attained within a multi-objective optimization framework. Upon completion of the training, the mesh generator has become capable of producing meshes that provide converged solutions within desired computational costs for new configurations in a single simulation, eliminating the need for iterative CFD processes typically required for grid convergence.

    To mitigate the substantial computational expense associated with data acquisition using CFD simulations, the training process has been divided into two steps. Initially, meshing parameters that affect the shape of a mesh have been optimized to maximize geometric mesh quality, assessed by the ratio of determinants of the Jacobian matrices and skewness. Subsequently, resolution-related meshing parameters have been optimized by integrating CFD simulations. Furthermore, multi-agent reinforcement learning has been implemented, with 256 agents performing CFD simulations simultaneously, to accelerate the data acquisition process.
    
    The effectiveness of the developed method has been assessed through mesh generation across a broad range of blade passage configurations. The trained network has been identified for managing diverse geometric profiles, including high-pressure and low-pressure turbine blades, axial compressor blades, and impulse rotor blades. In addition, the optimal mesh resolution can be determined prior to the simulation according to flow conditions, including intricate physical phenomena such as boundary layers, shock waves, and flow separation. The optimality of the method has been confirmed by comparing the accuracy and efficiency obtained in a single attempt by the current method with those derived from the conventional iterative optimization approach.

    The present method exhibits an outstanding performance in automating the CFD process for a blade passage configuration in mesh generation. Extending the proposed approach to accommodate configurations beyond blade profiles is contemplated as a prospective future direction. Dynamic optimization of numerical schemes and physical models is another future direction of the current research, aiming for the complete automation of the CFD process.
    
    \section*{CRediT authorship contribution statement}
    \textbf{Innyoung Kim:} Conceptualization, Investigation, Methodology, Software, Validation, Visualization, Writing - Original Draft. \textbf{Jonghyun Chae:} Investigation, Visualization. \textbf{Donghyun You:} Conceptualization, Funding Acquisition, Supervision, Writing - Original Draft.
	
    \section*{Declaration of competing interest}
    The authors declare that they have no known competing financial interests or personal relationships that could have appeared to influence the work reported in this paper.
    
    \section*{Acknowledgements}
    The work was supported by the National Research Foundation of Korea (NRF) under the Grant Number NRF-2021R1A2C2092146 and RS-2023-00282764 and the Korea Institute of Energy Technology Evaluation and Planning (KETEP) under Grant Number RS-2023-00242282.
 
    \newpage
    \bibliographystyle{elsarticle-num}
    \biboptions{sort&compress}
    \bibliography{DRL_Mesh_CFD}
	
    \newpage
    \listofalgorithms
	
    \pagebreak
    \clearpage
    {\fontsize{10}{10}\selectfont
    \begin{algorithm}[H]
    {
        \label{alg_drl_1st}
	    \caption{Single-step actor-critic method for training a mesh generator for geometric quality maximization.}
		initialize actor network $\bf{\pi}_{\bf{\phi}}$ and critic network $Q_{\bf{\zeta}}$ with random parameters $\bf{\phi}$, $\bf{\zeta}$;\\
		initialize episode and replay buffer $\mathcal{B}$;\\
		\Repeat{convergence}{
		    episode $\gets$ episode+1;\\
    		randomly select state $\bf{s}$;\\
    		determine action with exploration noise: $\bf{a} \gets \bf{\pi}_{\bf{\phi}}(\bf{s})+\bf{\epsilon}$;\\
    		generate mesh and evaluate reward $r$;\\
            store data of $(\bf{s},\bf{a},r)$ in $\mathcal{B}$;\\

        	sample mini-batch of $N_b$ data from $\mathcal{B}$;\\
        	update $\bf{\zeta}$ with the loss $N_b^{-1}\sum(r-Q_{\bf{\zeta}}(\bf{s},\bf{a}))^2$;\\			
        	\If{\textnormal{episode} \textup{mod} $2$}{
            	update $\bf{\phi}$ by the deterministic policy gradient $N_b^{-1}\sum\nabla_{\bf{a}} Q_{\bf{\zeta}}(\bf{s},\bf{a})|_{\bf{a}=\bf{\pi}_{\bf{\phi}}(\bf{s})}\nabla_{\bf{\phi}}\bf{\pi}_{\bf{\phi}}(\bf{s})$;\\
        		}	
		}
	}
    \end{algorithm}
    }

    \pagebreak
    \clearpage
    {\fontsize{10}{10}\selectfont
    \begin{algorithm}[H]
    {
    \label{alg_drl_2nd}
	    \caption{Multi-agent single-step actor-critic method for training a mesh generator for optimal CFD.}
		initialize actor network $\pi_{\phi}$ and critic network $Q_{\theta}$ with random parameters $\phi$, $\theta$;\\
		initialize episode and replay buffer $\mathcal{B}$;\\
        \tcp{Initial task distribution}
        \For{\textup{agent} $i=1$ \KwTo $N_{agent}$}{
                randomly select state $\bf{s}_{i}$;\\
    		determine action with exploration noise: $\bf{a}_{i} \gets \bf{\pi}_{\bf{\phi}}(\bf{s}_{i})+\bf{\epsilon}$;\\
    		generate mesh and run CFD;\\
        }
        \tcp{Dynamic task distribution}
        \Repeat{convergence}{
            \If{\textup{CFD of agent $i$ is completed}}{
                compute error function $r_{e, i}$ and cost function $r_{c, i}$ from simulation;\\
                compute reward $r_{i}$ and reproduce data;\\
                store reproduced data of $(\bf{s}_{i},\bf{a}_{i},r_{i})_{1}, \dots ,(\bf{s}_{i},\bf{a}_{i},r_{i})_{N_{repro}}$ in $\mathcal{B}$;\\
                
                \tcp{Training loop}
                \For{$l=1$ \KwTo $l_{max}$}{
        		sample mini-batch of $N_b$ data from $\mathcal{B}$;\\
        		update $\theta$ with the loss $N_b^{-1}\sum(r-Q_{\theta}(\bf{s},\bf{a}))^2$;\\			
        		\If{$l$ \textup{mod} $2$}{
            		update $\phi$ by the deterministic policy gradient 
              $N_b^{-1}\sum\nabla_{\bf{a}} Q_{\theta}(\bf{s},\bf{a})|_{\bf{a}=\pi_{\phi}(\bf{s})}\nabla_{\phi}\pi_{\phi}(\bf{s})$;\\
        		}
        	}
                
                \tcp{Transition to a new episode}
                episode $\gets$ episode+1;\\
                randomly select new state $\bf{s}_{i}$;\\
    		determine action with exploration noise: $\bf{a}_{i} \gets \bf{\pi}_{\bf{\phi}}(\bf{s}_{i})+\bf{\epsilon}$;\\
    		generate mesh and run CFD;\\
            }
		}
	}
    \end{algorithm}
    }
    
    \newpage
    \listoftables
	
    \pagebreak
    \clearpage
    \begin{table}[h]
    \caption{Parameters required for the mesh generator and their descriptions. The initial five parameters construct the boundary shape of the mesh, while the subsequent parameters control the mesh resolution.}
    \begin{center}
    \begin{footnotesize}
    \begin{tabularx}{\textwidth}{ l l } 
        \hline
        Boundary construction & Description \\
        \hline
        $y_{in}$ & Inlet position in the vertical direction  \\
        $y_{out}$ & Outlet position in the vertical direction  \\
        $\alpha_{camber}$ & Degree of the curvature of the periodic boundaries\\
        & following the camber line  \\
        $x^{o}_{in}$ & HO-type interface position in the horizontal direction \\
        $x^{o}_{out}$ & OH-type interface position in the horizontal direction \\ \hline

        Resolution control & Description  \\
        \hline
        $N_{t}$ & Number of nodes of the O-type mesh\\ 
        & in the tangential direction to the blade surface  \\
        $N_{n}$ & Number of nodes of the O-type mesh\\
        & in the normal direction to the blade surface  \\
        $\beta_{le}$ & Degree of clustering of nodes at the leading edge  \\
        $\beta_{te}$ & Degree of clustering of nodes at the trailing edge  \\
        $\Delta n_{1}$ & First cell height in the direction normal to the blade surface \\
        \hline
    \end{tabularx}
    \end{footnotesize}
    \end{center}
    \label{table_params}
    \end{table}

    \newpage
    \listoffigures
	
    \pagebreak
    \clearpage

    \pagebreak
    \clearpage

    \pagebreak
    \clearpage

    \pagebreak
    \clearpage

    \pagebreak
    \clearpage

    \pagebreak
    \clearpage

    \pagebreak
    \clearpage

    \pagebreak
    \clearpage

    \pagebreak
    \clearpage

    \pagebreak
    \clearpage

    \pagebreak
    \clearpage

    \pagebreak
    \clearpage

    \pagebreak
    \clearpage

    \pagebreak
    \clearpage

    \pagebreak
    \clearpage

    \pagebreak
    \clearpage

    \pagebreak
    \clearpage

    \pagebreak
    \clearpage

    \pagebreak
    \clearpage

    \pagebreak
    \clearpage

    \pagebreak
    \clearpage

    \pagebreak
    \clearpage

    \pagebreak
    \clearpage

    \pagebreak
    \clearpage
    
    \appendix
    \section{Effect of the size of the neural network} \label{app:network_size}
    To investigate whether neural networks with four hidden layers (512, 256, 256, and 128 neurons) exhibit sufficient performance for the current optimization problem, networks with varying numbers of hidden layers and neurons are trained. Additional three sets of networks are configured, each with both the actor and critic networks having identical sizes: one layer (128 neurons), two layers (256 and 128 neurons), and six layers (512, 512, 256, 256, 128, and 128 neurons). For a direct comparison of the impact of network size, and considering the substantial computational costs involved in data acquisition (eight weeks), learning is conducted in an offline reinforcement learning setup. In this setup, the training of the networks is conducted on a fixed dataset previously obtained through $5\times10^5$ RANS simulations. For comparison, the normalized cost function $r_c^*$ and the normalized error function $r_e^*$ during the learning process are tracked as a function of the number of episodes. The LS89 cascade blade at the MUR43 and MUR47 test conditions, as analyzed in Sections~\ref{sec5.1} and \ref{sec5.4}, is examined for $w^{*}$ values of 0 and 1.

\begin{figure}[]
    \centering
    \begin{subfigure}[t]{0.495\linewidth}
    \includegraphics[width=1.0\linewidth]{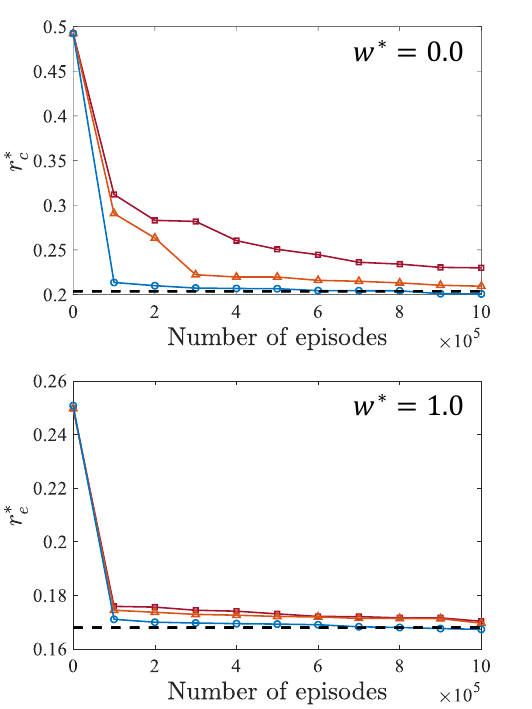}
    \caption{} \label{fig_network_size_sub}
    \end{subfigure}
    \begin{subfigure}[t]{0.495\linewidth}
    \includegraphics[width=1.0\linewidth]{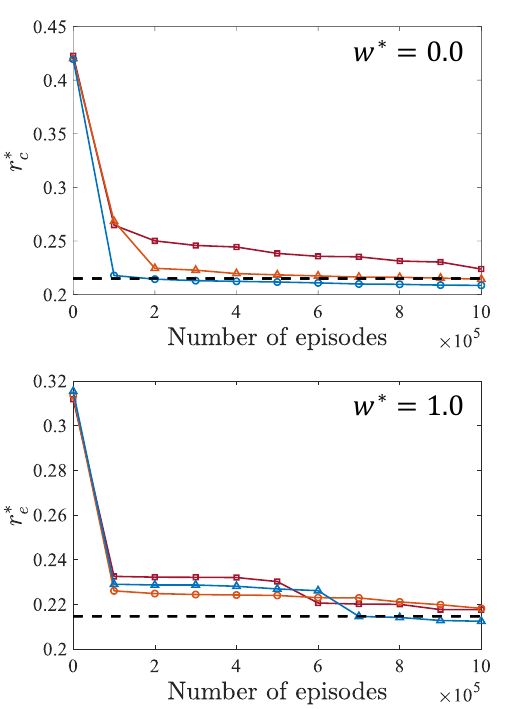}
    \caption{} \label{fig_network_size_super}
    \end{subfigure}
    \caption{The normalized cost function $r_c^*$ and the normalized error function $r_e^*$ as a function of the number of episodes for the LS89 cascade blade under (a) the MUR43 and (b) MUR47 test cases for $w^{*}$ values of 0 and 1. \colr{$\square$}, one layer (128 neurons); \colo{$\triangle$}, two layers (256 and 128 neurons); \colb{$\bigcirc$}, six layers (512, 512, 256, 256, 128, and 128 neurons). The results obtained from the present network of four layers (512, 256, 256, and 128 neurons) are depicted with dashed lines for comparison.} \label{fig_network_size}
    \end{figure}

    As illustrated in Fig.~\ref{fig_network_size}, convergence is observed at $10^6$ episodes for all cases. An increase in the size of the network results in improved performance. However, it is observed that the marginal gains in performance due to an increase in network size gradually decrease. Consequently, the configuration with four hidden layers is found to be sufficient for the current optimization problem, with a performance difference of at most 4\% compared to the six-hidden-layer configuration.

    \pagebreak
    \clearpage
    
\pagebreak
\clearpage

\end{document}